\documentclass[twocolumn,prb,showpacs,preprintnumbers,amsmath,amssymb,superscriptaddress,groupedaddress,reprint]{revtex4-1}[11pt]

\usepackage[pdftex]{graphicx}
\usepackage{epstopdf}
\usepackage{amsmath}
\usepackage{amsfonts}
\usepackage{amssymb}
\usepackage{subfigure}
\usepackage{color}
\usepackage{float}
\usepackage{color}
\usepackage{epstopdf}

\begin{document}
\title{Ultra-sensitive nanoscale magnetic field sensors based on resonant spin filtering}
\author{Abhishek Sharma}
\author{Ashwin. A. Tulapurkar}
\author{Bhaskaran Muralidharan}
\affiliation{Department of Electrical Engineering, Indian Institute of Technology Bombay, Powai, Mumbai-400076, India}
\date{\today}
\medskip
\widetext
\begin{abstract}
Solid state magnetic field sensors based on magneto-resistance modulation find direct applications in communication devices, specifically in proximity detection, rotational reference detection and current sensing. In this work, we propose sensor structures based on the magneto-resistance physics of resonant spin-filtering and present device designs catered toward exceptional magnetic field sensing capabilities. Using the non-equilibrium Green's function spin transport formalism self consistently coupled to the Poisson's equation, we present highly-tunable pentalayer magnetic tunnel junction structures that are capable of exhibiting an ultra-high peak tunnel magneto resistance $(\approx 2500 \%$).  We show how this translates to device designs featuring an ultra-high current sensitivity enhancement of over 300\% in comparison with typical trilayer MTJ sensors, and a wider tunable range of field sensitivity. We also demonstrate that a dynamic variation in sensor functionalities with the structural landscape enables a superior design flexibility over typical trilayer sensors. An optimal design exhibiting close to a 700\% sensitivity increase as a result of angle dependent spin filtering is then presented.This work sets a stage to engineer spintronic building blocks via the design of functional structures tailored to exhibit ultra-sensitive spin filtering.
\end{abstract}
\pacs{}
\maketitle
\section{Introduction}
The emerging area of spintronics relies on the storage, control and manipulation of spin information via magnets and spin currents \cite{bader}.  Devices built using spintronics include memory cells \cite{Grollier}, switches \cite{Nitta}, oscillators \cite{kat,Osc_Kim,parkin,bader_2,Sanchar}, rectifiers \cite{Tulapurk,Miwa}, magnetic field sensors \cite{STapplications,zeng2012,VanDijken2005} and interconnects \cite{bader,behtash}, to name a few. While the basic entity of information processing is the state of the magnet and may be manipulated, in principle, via spins alone \cite{behtash},  it is electrical read-write processes \cite{bader,bader_2} that are of immediate technological consequence. \\
\indent In the context of magnetic tunnel junctions (MTJ), which are actively researched spintronic building blocks, the sensitivity of an electrical read-write process depends on the difference between the resistances of relative magnet orientations, and is quantified by the tunnel magneto resistance (TMR) defined as 
\begin{equation}
TMR=\frac{R_{AP}-R_{P}}{R_{P}},
\end{equation}
where $R_{P}$ and $R_{AP}$ represent device resistances when the relative orientation between the fixed and free magnetic layers are parallel and anti-parallel respectively. A high TMR is usually desired and may be typically enhanced via the physics of spin-filtering \cite{butler,berger,slon,bauer,Ralph1,Butler2001}. A typical trilayer device has a peak TMR in the order of $\approx 200 \%$ \cite{suzuki,kubota} and is not easily tunable to higher values due to the limited design landscape imposed by the physics of single barrier tunneling \cite{chatterji}.\\
\begin{figure}[htb!]
	\centering
	\includegraphics[width=3.2in]{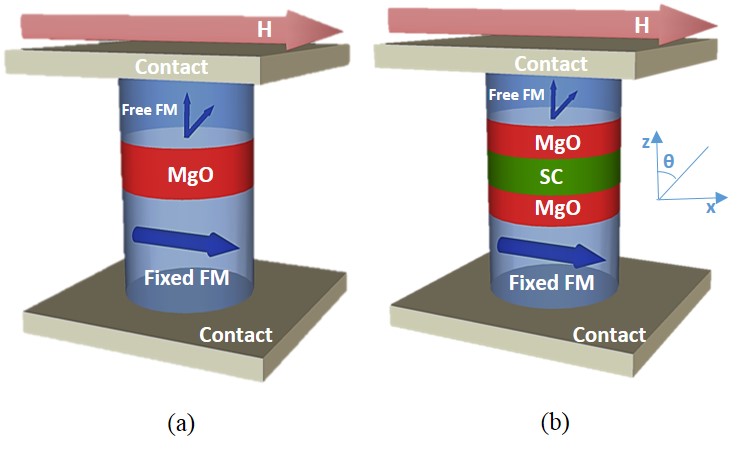}
	\caption{Sensor device prototype.The trilayer device comprises (a) a typical MTJ with an insulating MgO layer between the fixed and free ferromagnetic layers. (b) An RTMTJ based device comprises a MgO-Semiconductor-MgO heterostructure between the fixed and free layers. An external magnetic field ($H$) is applied along the $\hat{x}$ direction of the free layer.}
	\label{device_schematic}
\end{figure}
\begin{figure}[htb!]
	\centering
	\includegraphics[width=3.5in,height=1.6in]{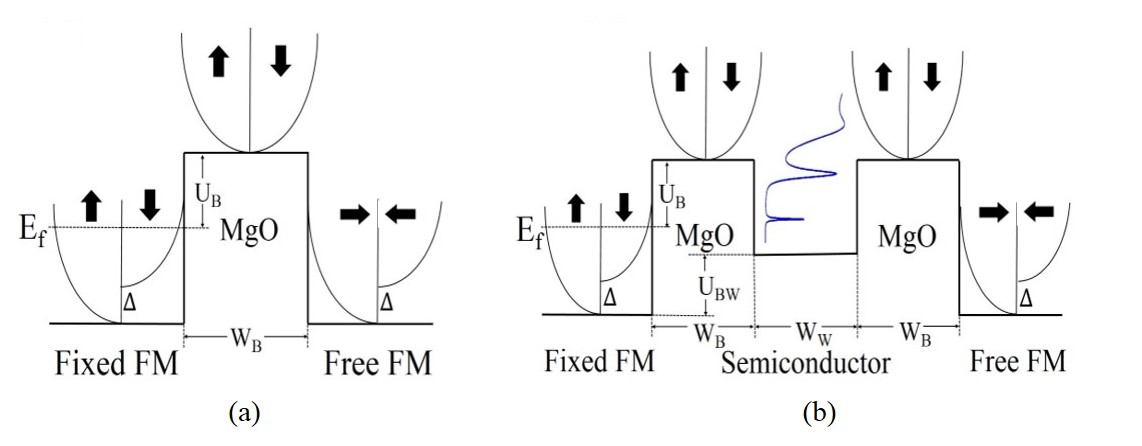}
	\caption{Energy band schematic. (a) A trilayer MTJ sensor device at equilibrium along $\hat{z}$ direction. The ferromagnetic contacts have an exchange energy of $\Delta$ with $E_f$ being Fermi energy, $U_B$, the barrier height in MgO above Fermi Energy. (b) An RTMTJ device at equilibrium along $\hat{z}$ directoin. Here, $U_{BW}$ is the difference between the bottom of the conduction band of the ferromagnet and the semiconductor. The resonant peaks shown in the inset are responsible for the ultra-sensitive spin filtering to be discussed here.}
	\label{device_schematic_2}
\end{figure}
\indent In this article, we attempt to alleviate some of the above mentioned issues by proposing the use of a {\it{resonant tunneling}} magnetic tunnel junction (RTMTJ) structure \cite{chatterji,RMTJnanolatter2008,Zheng1999} which aims to harness the sensitive spin filtering capabilities enabled via double barrier tunneling. We demonstrate that such an enhanced filtering leads to an ultra-high peak TMR ($\approx 2500 \%$), whose value may in turn be tuned via appropriate structural design. The possibility of a high peak TMR value resulting from the physics of double barrier resonant tunneling has been hinted at theoretically \cite{asymm_rtd}, and observed experimentally \cite{RMTJnanolatter2008} in related structures. An earlier theoretical work \cite{Bauer_TSTT} also established the role of oxygen vacancies of the MgO barrier in a trilayer structure on the resonant enhancement of thermal spin torques in trilayer MTJ structures. While the ultra-high TMR presents several practical applications in sight, we explore one that is ubiquitous in miniaturized communication devices, namely, magnetic field sensing and present magneto-resistive (MR) structures aimed at displaying exceptional magnetic field sensor capabilities. \\
\indent Currently explored paradigms in solid state magnetoresistance (MR) sensing typically employ a trilayer MTJ \cite{zeng2012,VanDijken2005,Hehn2004,Rijks1994} structure, as sketched in the schematic in Fig.~\ref{device_schematic}(a), with an initial alignment of the fixed and free magnets being perpendicular to each other. The sensor functions as follows: the device is kept under the constant voltage and with the application of a magnetic field, the magnetization of the free layer settles at a new equilibrium angle leading to a change in the current as the magnetization angle between the fixed and free layer varies, which typically increases with the tunnel magneto resistance (TMR) of the junction. \\
\indent The proposed RTMTJ based device structure is depicted schematically in Fig.~\ref{device_schematic}(b), which comprises a heterostructure quantum well sandwiched between the fixed and free layers. The RTMTJ structure may be realized by means of semiconductors such as ZnO\cite{ZnO_hetro}, Ge\cite{Shi2012}, GaAs\cite{Lu2006}, AlN\cite{Yang2009}, InN\cite{Zhang2008} and others.  With our proposed structure capable of exhibiting an ultra-high TMR, we present device designs featuring an ultra-high current sensitivity enhancement of over 300\% in comparison with typical trilayer MTJ sensors, and a wider tunable range of field sensitivity. We also demonstrate that a dynamic variation in sensor functionalities with the structural landscape enables a superior design flexibility over typical trilayer sensors. As an important corollary, we show angle dependent spin-filtering can further optimize the device design resulting in $\approx 700\%$ sensitivity increase. \\
\indent With the above mentioned groundwork, we reinforce the double barrier spin filtering physics rigorously by employing the non-equilibrium Green's function (NEGF) spin transport formalism coupled with the Poisson's equation. The structures shown schematically in Fig.~\ref{device_schematic} (a) and (b) are simulated using the parameterized tight binding NEGF framework described in previous works \cite{butler,deepanjan,yanik,datta2,lunds,akshay}. Using this formulation sketched briefly in Section II, we turn to our device design process in Section III B by first elucidating the physics of double barrier spin filtering and how it translates to an ultra-high TMR. Following this, in Section III C, we consider two RTMTJ device designs and demonstrate their superior performance as magnetic field sensors in comparison with trilayer designs by examining the sensitivity, nonlinearity and linear sensing range. We also depict in Section III D, trends on how the sensor performance may be further tuned via the penta-layer device design landscape to establish the superior design flexibility that our proposal carries. In particular, in Section III E, the optimal design as a result of angle dependent spin filtering is presented. 
\section{Theoretical Formulation}
In this section, we sketch the essential details of the NEGF simulation procedure \cite{butler,yanik,deepanjan,lunds,akshay} that will be used to analyze the sensor device designs, based on the device structures detailed in Fig.~\ref{device_schematic_2}. The trilayer MTJ has a layer of MgO between the magnets while the RTMTJ has a heterostructure of MgO-Semiconductor-MgO sandwiched between the fixed and the free magnets leading to resonant peaks in the transmission spectrum. The devices are held at a fixed voltage during detection of the magnetic field, small enough such that the resulting spin current does not excite significant magnetization dynamics. The magnetization of the fixed layer is along the $\hat{x}$-axis in both cases and that of the free layer at zero field is along the $\hat{z}$ direction \cite{zeng2012,VanDijken2005,Hehn2004,Rijks1994}. The applied magnetic field to be sensed is along the $\hat{x}$ direction of the free layer.  \\
\indent The NEGF formalism solved self-consistently with the Poisson's equation within the effective mass framework is employed to calculate the charge currents in the devices \cite{datta2,butler,lunds,akshay,yanik}. We start with the energy resolved spin dependent single particle Green's function matrix $[G(E)]$ evaluated from the device Hamiltonian matrix $[H]$ given by:
\begin{equation}
[G(E)] = [EI-H-\Sigma_T-\Sigma_B]^{-1},
\end{equation}
where the device Hamiltonian matrix,  $[H]=[H_0]+[U]$, comprises the device tight-binding matrix, $[H_0]$ and the Coulomb charging matrix ,$[U]$, in real space, $[I]$ is the identity matrix with the dimensionality of the device Hamiltonian. The quantities $[\Sigma_T]$ and $[\Sigma_B]$ represent the self-energy matrices \cite{datta2} of the top and bottom magnetic layers evaluated within the tight-binding framework \cite{yanik,deepanjan}.  A typical matrix representation of any quantity $[A]$ defined above entails the use of the matrix element $A(z,z',k_x,k_x',k_y,k_y',E)$, indexed on the real space $z$ and the transverse mode space $k_x,k_y$. To account for the finite cross-section, we follow the uncoupled transverse mode approach, with each transverse mode indexed as $k_x,k_y$ evaluated by solving the sub-band eigenvalue problem \cite{lunds,salah2,datta3}. \\
\indent The charging matrix, $[U]$, is obtained via a self consistent calculation with the Poisson's equation along the transport direction $\hat{z}$ given by
\begin{eqnarray}
\frac{d}{dz}\left(\epsilon_r(z)\frac{d}{dz} U(z)\right)=\frac{-q^2}{\epsilon_0}n(z) \label{poisson}\\
n(z)=\frac{1}{A.a_0}\displaystyle\sum_{k_x,k_y}G^n(z;k_x,k_y),
\label{n_r}
\end{eqnarray}
with $G^n(z;k_x,k_y)=G^n(z,z,k_x,k_x,k_y,k_y)$, being a diagonal element of the energy resolved electron correlation matrix $[G^n(E)]$ given by
\begin{eqnarray*}
[G^n(E)]= \int dE [G(E)] [\Gamma_T(E)] [G(E)]^{\dagger} f_T(E)  \\ \nonumber
\qquad + [G(E)][\Gamma_B(E)] [G(E)]^{\dagger} f_B(E). \\
\label{G_n}
\end{eqnarray*}
Here, $[\Gamma_T(E)]=i\left ([\Sigma_T(E)]- [\Sigma_T(E)]^{\dagger} \right )$ and $[\Gamma_B(E)]=i\left( [\Sigma_B(E)]-[\Sigma_B(E)]^{\dagger} \right )$ are the spin dependent broadening matrices \cite{datta2} of the top and bottom contacts. The Fermi-Dirac distributions of the top and bottom contacts are given by $f_T(E)$ and $f_B(E)$ respectively. Here, $U(z)$ is the potential profile inside the device subject to the boundary conditions, $U_{FixedFM}=-qV/2$ and $U_{FreeFM}=qV/2$, with $V$ being the applied voltage, $A$ being the cross sectional area of the device, $a_0$ being the inter-atomic spacing in effective mass framework and $\hbar$ being the reduced Planck's constant. \\
\indent The summit of the calculation is the evaluation of charge currents following the self-consistent convergence of \eqref{poisson} and \eqref{n_r}. The matrix element of the charge current operator $\hat{I}_{op}$ representing the charge current between two lattice points $i$ and $i+1$ is given by \cite{datta1}
\begin{equation}
{I}_{op,i,i}=\frac{i}{\hbar}\left(H_{i,i+1}G^{n}_{i+1,i}-G^{n\dagger}_{i,i+1}H^{\dagger}_{i+1,i}\right) ,
\end{equation}
following which the charge current $I$ is given by  $I =q \int dE \text{ Real [Trace(}\hat{I}_{op}\text{)]}$, where, the current operator $\hat{I}_{op}$ is a 2$\times$2 matrix in spin space, $H$ is the Hamiltonian matrix of the system and $q$ is the electronic charge. 
\indent We use the Landau-Lifshitz-Gilbert (LLG) equation to calculate the equilibrium magnetization of the free layer in the presence of an applied magnetic field \cite{slon,brat}:
\begin{equation}
\left( 1+\alpha^{2}\right) \frac{\partial \hat{m}}{\partial t} = -\gamma \hat{m} \times \vec{H}_{eff} - \gamma \alpha \left( \hat{m} \times ( \hat{m} \times \vec{H}_{eff})\right) 
\nonumber
\end{equation}
where $\hat{m}$ is the unit vector along the direction of magnetization of the free magnet, $\gamma$ is the gyromagnetic ratio of the electron, $\alpha$ is the Gilbert damping parameter, $\vec{H}_{eff} = \vec{H}_{app} + H_km_{z}\hat{z}$ is the effective magnetic field with $\vec{H}_{app}$ being the applied external field, and $H_k$ being the anisotropy field. \\
\begin{figure}[!thb]
	\subfigure[]{\includegraphics[scale=0.21]{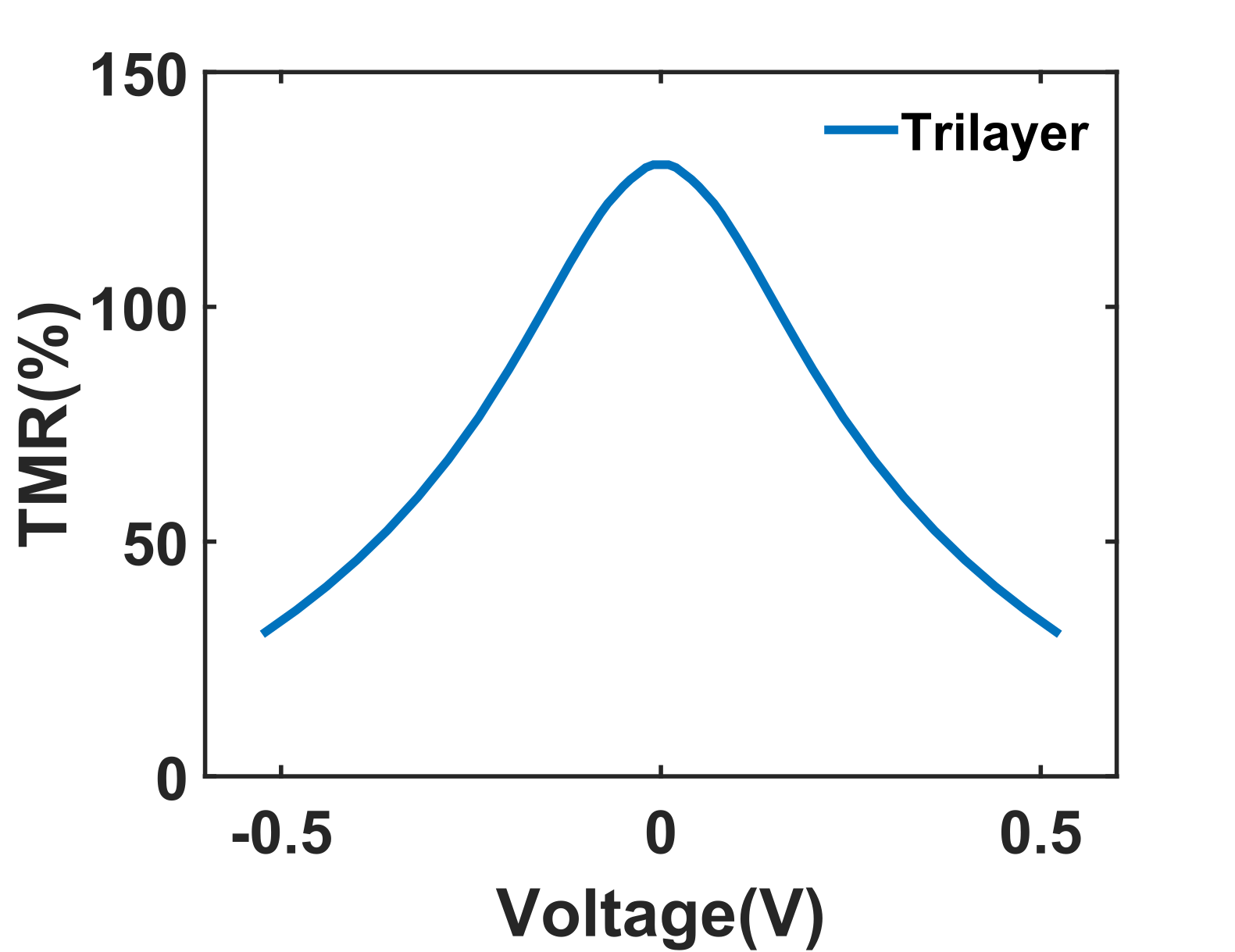}}
	\subfigure[]{\includegraphics[scale=0.21]{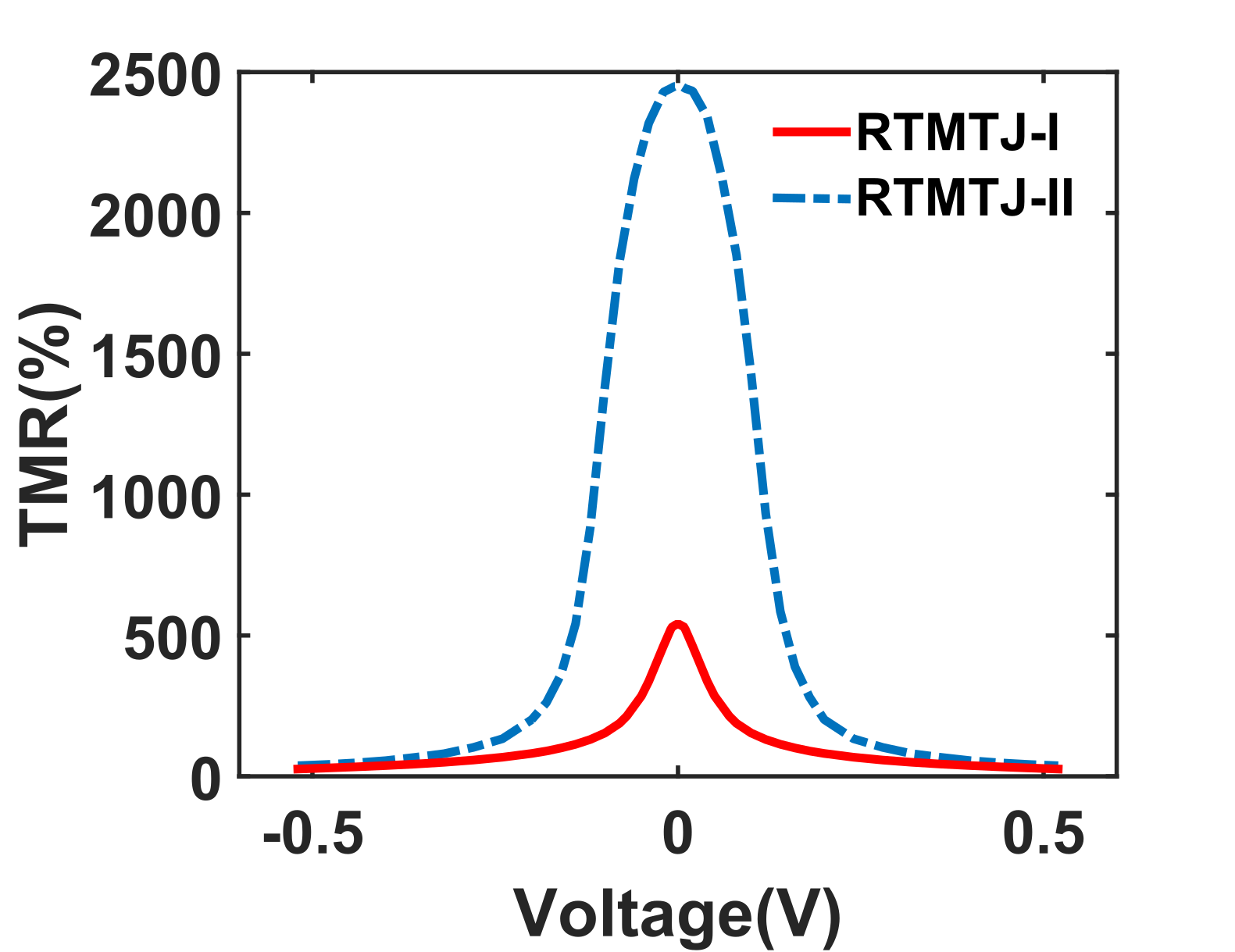}}
	\caption{Device TMR profiles. (a) The TMR variation with voltage for a trilayer MTJ device, (b) for an RTMTJ device. In case of (b), the TMR of the device can be varied to a large extent via the proper positioning of the resonant peaks schematically depicted in Fig.~\ref{device_schematic}(d). As a sample, we plot the TMR profiles of two such device designs (see text) labeled I and II, which are to be considered in detail.}
	\label{TMR_V}
\end{figure}
\begin{figure}
	\centering
	\includegraphics[width=3.5in]{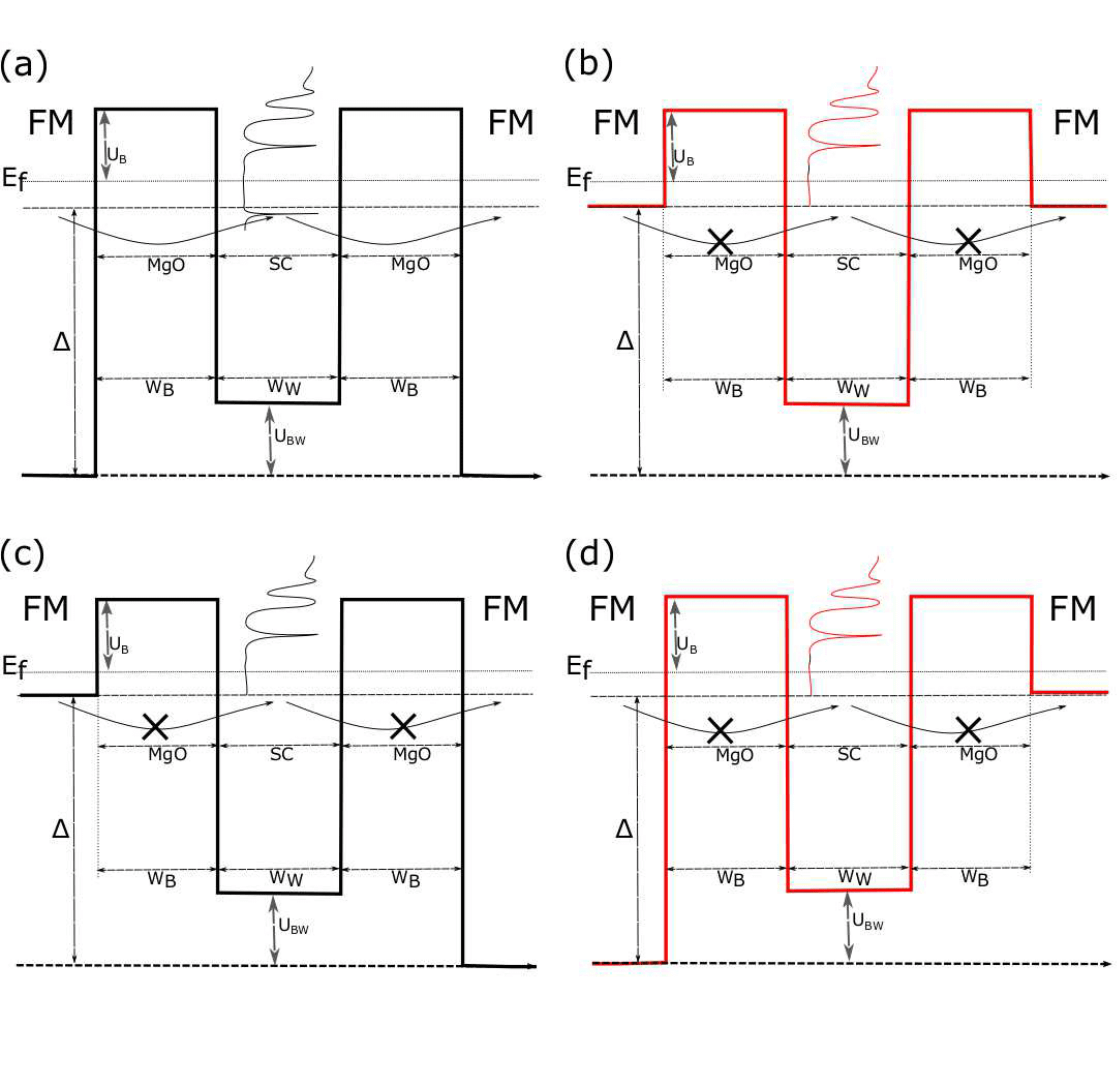}
	\caption{Spin filtering physics. The spin dependent energy band schematic of the RTMTJ device for the (a,b) parallel configuration, and for the (c,d) anti-parallel configuration. A schematic of the transmission function in each case is also shown. (a) Band profile seen by up-spin electrons (black) in the parallel configuration. The up-spin transmission peaks can occur just below $\Delta$. (b) Band profile seen by the down-spin electrons (red) in the parallel configuration. (c) Band profile seen by the up-spin electrons (black) in the anti-parallel configuration. (d) Band profile seen by the down-spin electrons (red) in the anti-parallel configuration.}
	\label{band_diagram}
\end{figure}
\begin{figure}[htb!]
	\subfigure[]{\includegraphics[width=1.75in]{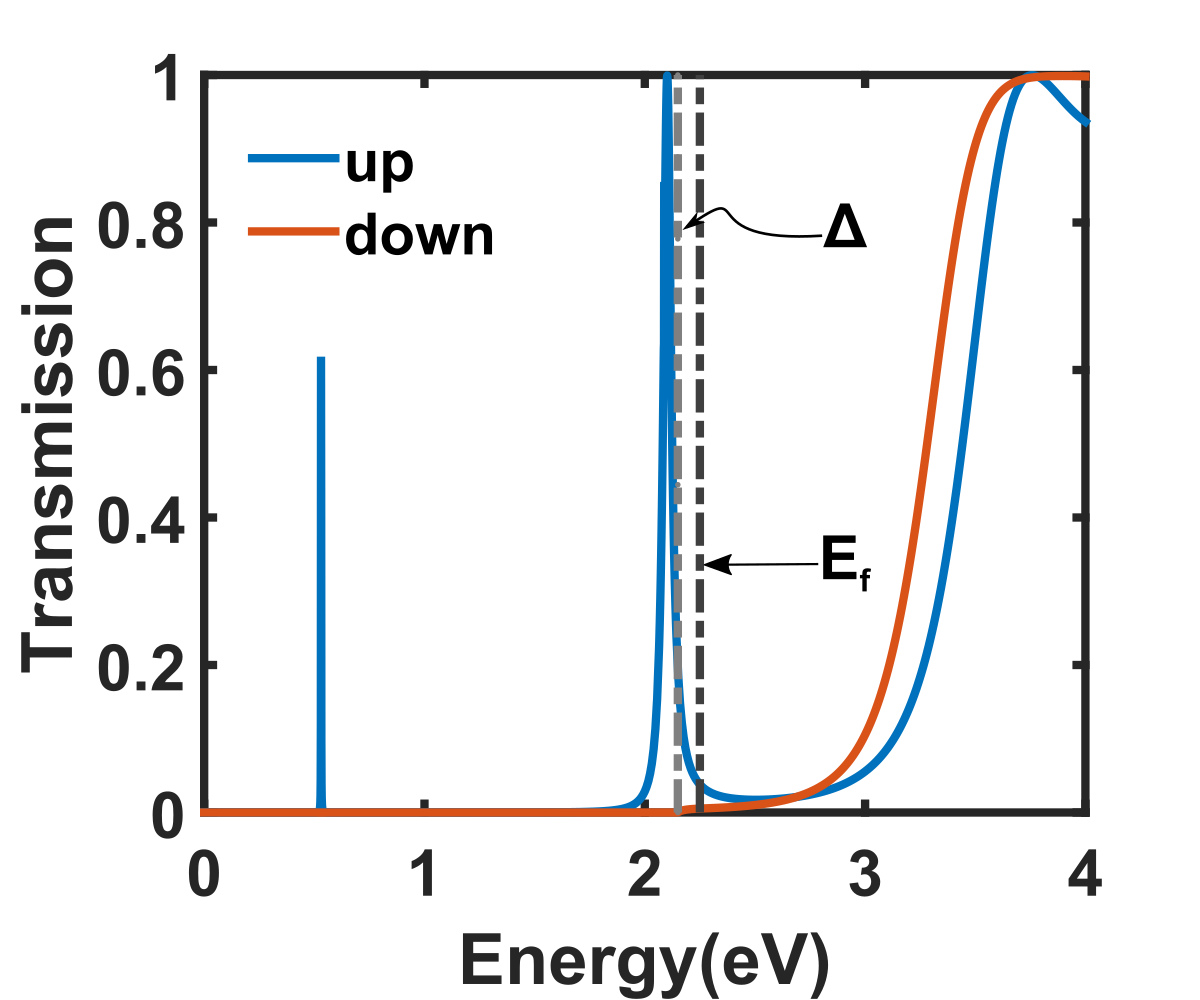}		}\subfigure[]{\includegraphics[width=1.75in]{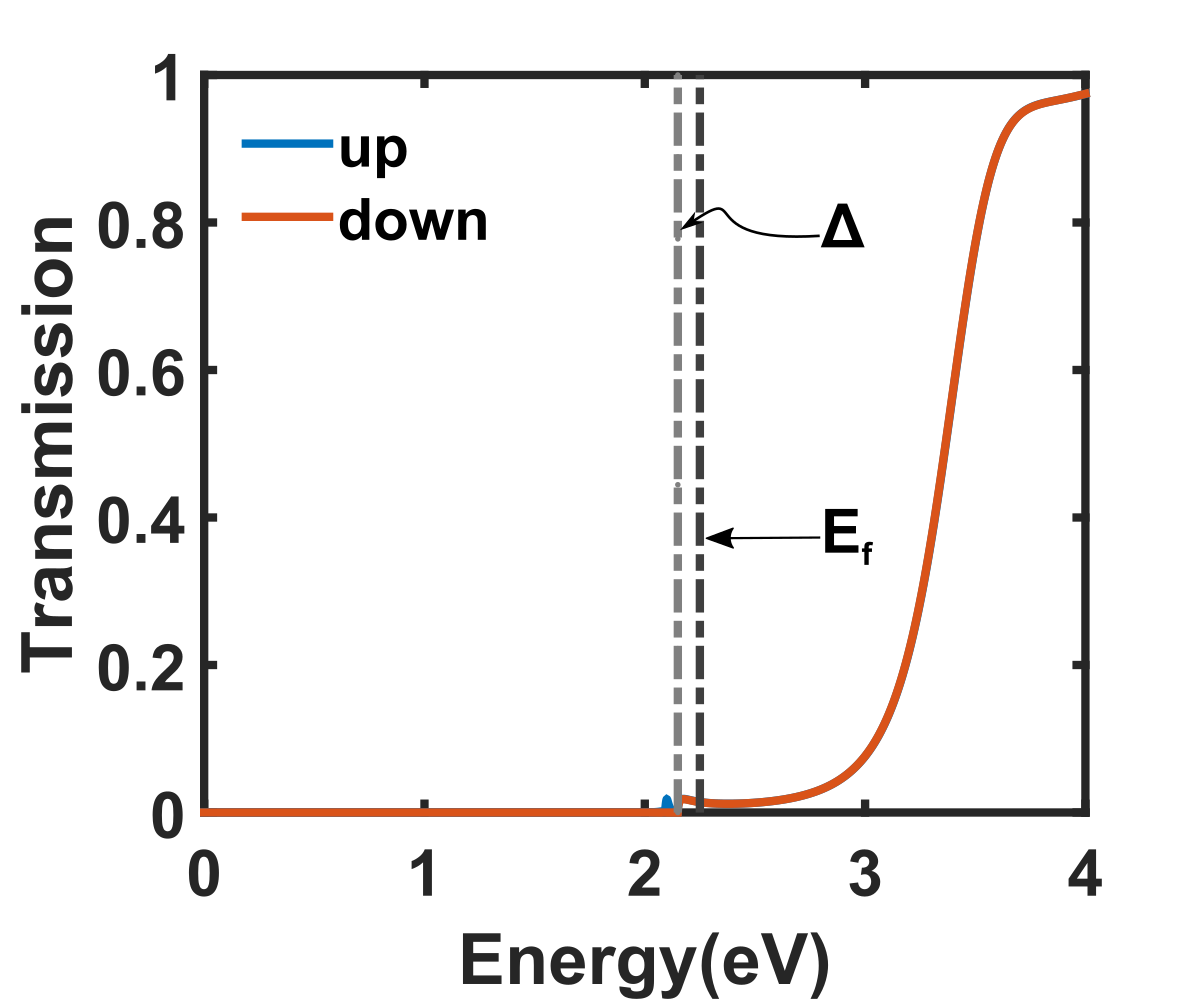}}
	\subfigure[]{\includegraphics[width=1.75in]{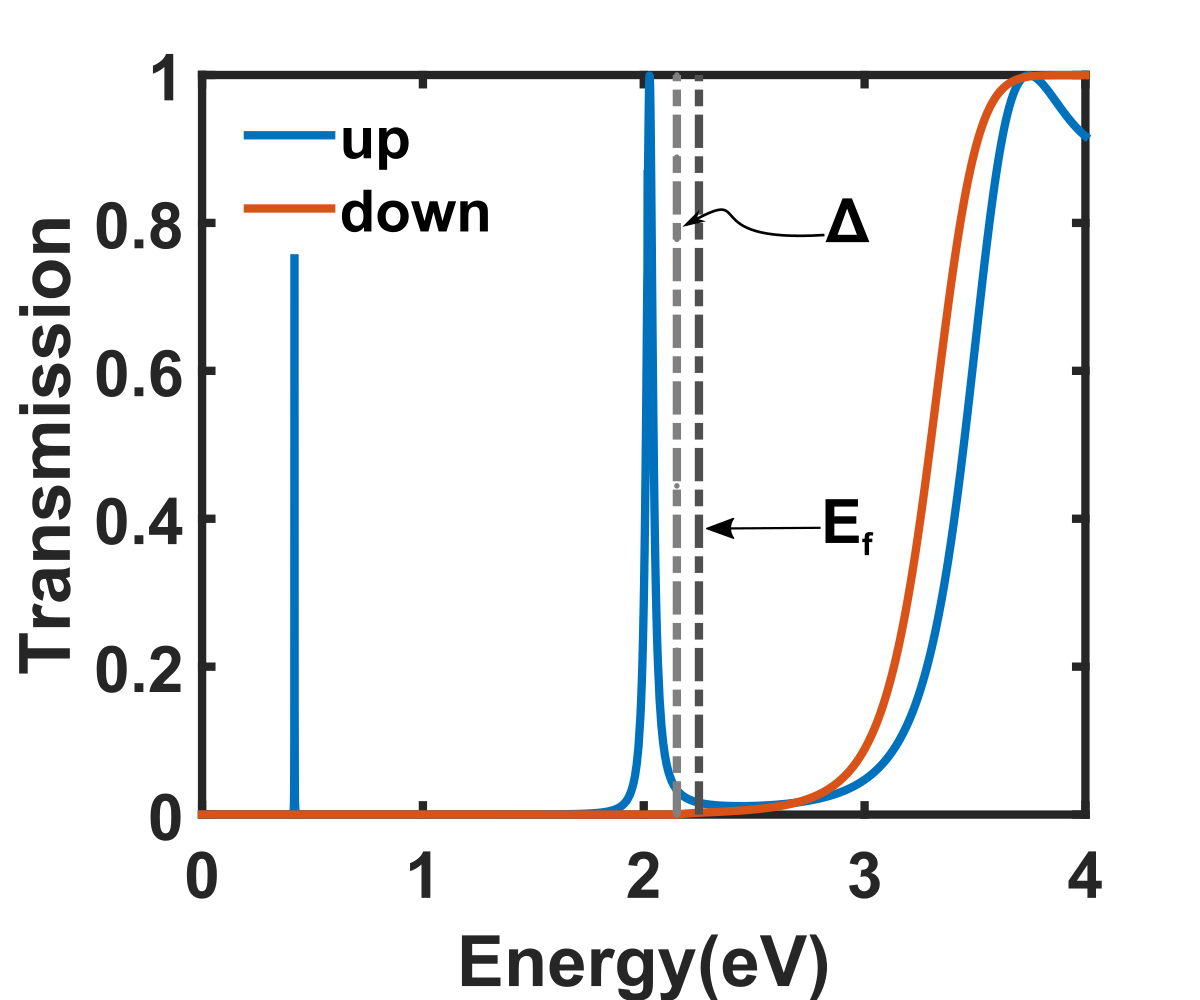}	}\subfigure[]{\includegraphics[width=1.75in]{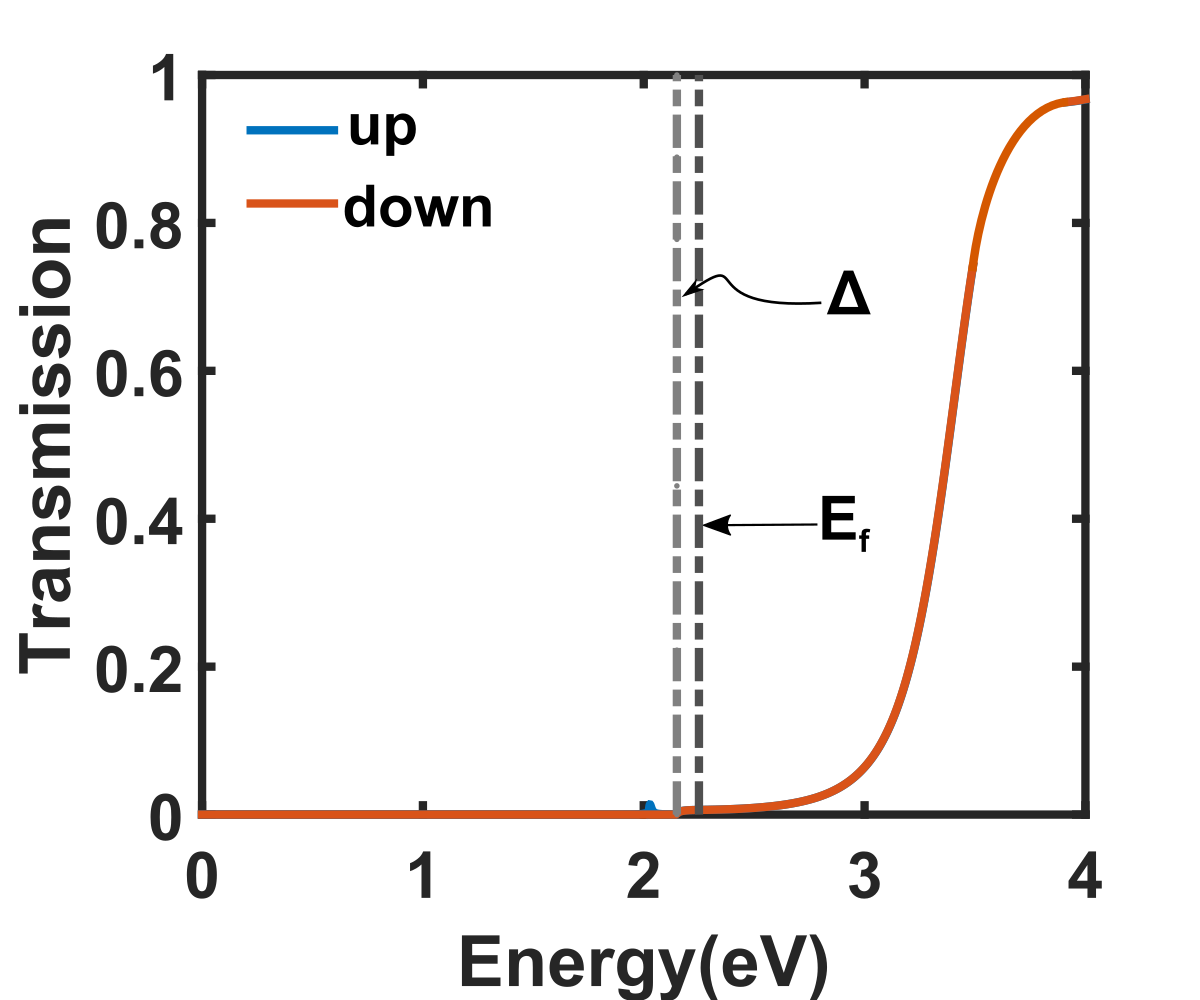}}
	\caption{Role of spin resolved transmission. Transmission peaks for RTMTJ-I at V=0. (a) Parallel configuration (b) Anti-parallel configuration. Transmission peaks for RTMTJ-II at V=0. (c) In the parallel configuration, only the up-spin transmission peak exists below $\Delta$. (d) In the anti-parallel configuration, there are no transmission peaks around the Fermi level resulting in a very small conductance.}
	\label{T_spectra}
\end{figure}
\section{Results and Discussion}
\subsection{Device structures}
\indent In our simulations, we use CoFeB as the ferromagnet with Fermi energy, $E_f = 2.25$eV and exchange splitting $\Delta = 2.15$ eV. The effective mass of MgO is $m_{OX} = 0.18 \ m_e$ and of the semiconductor, $m_{SC} = 0.36 \ me$, with $m_e$ being the free electron mass. The barrier height of the CoFeB-MgO interface is $U_B = 0.76$ eV above the Fermi energy \cite{datta3,kubota}. \\
\indent In the results that follow, the parameters chosen for the magnetization dynamics are $\alpha$ = 0.01, $\gamma$ = 17.6 MHz/Oe, with the anisotropy field $H_k$ varied over a range of $400-1650$Oe which translates to different thicknesses of the free layer in the real structure\cite{zeng2012,VanDijken2005}. The cross-sectional area of all devices considered is 70 $\times$ 160 nm\textsuperscript{2}. \\
\indent Sensor designs are evaluated based on an operating TMR. The TMR characteristics, as a function of the applied voltage of the trilayer and RTMTJ devices are shown in Fig.~\ref{TMR_V} (a) and (b) respectively. Specifically, we consider three device designs, namely, (i) the trilayer device (Fig.~\ref{TMR_V}(a)), with $W_B=1$nm (ii) RTMTJ device I  (bold in Fig.~\ref{TMR_V}(b)), with $W_B=1$nm, $W_{SC}=1$nm and $U_{BW}=-0.25$eV, having a  lower TMR ($TMR\approx500\%$ ) and (iii) RTMTJ device II (shown dotted Fig.~\ref{TMR_V}(b)), with $W_B=1$nm, $W_{SC}=1$nm and $U_{BW}=-0.45$eV, having an ultra-high TMR design ($TMR \approx 2500 \%$).\\ 
\subsection{Physics of spin filtering}
We now delve in to the TMR physics resulting via resonant enhancement of spin filtering on pentalayer structures, using equilibrium band diagrams depicted in
Fig.~\ref{band_diagram} for the first transverse mode. From Fig.~\ref{band_diagram}(a) and (b), it can be inferred that for the parallel configuration, the down-spin electron can not have a transmission peak below the exchange splitting $\Delta$, whereas no such restriction exists for the up-spin electrons. Therefore, a structure possessing up-spin transmission peaks just below $\Delta$ can also be designed.  This opens up an extra channel for the up-spin electrons in the parallel configuration while blocking the down-spin electrons depending upon the relative position of the ferromagnetic Fermi level with respect to the exchange splitting $\Delta$. \\
\begin{figure}[tb!]
	\subfigure[]{\includegraphics[width=1.7in,height=1.25in]{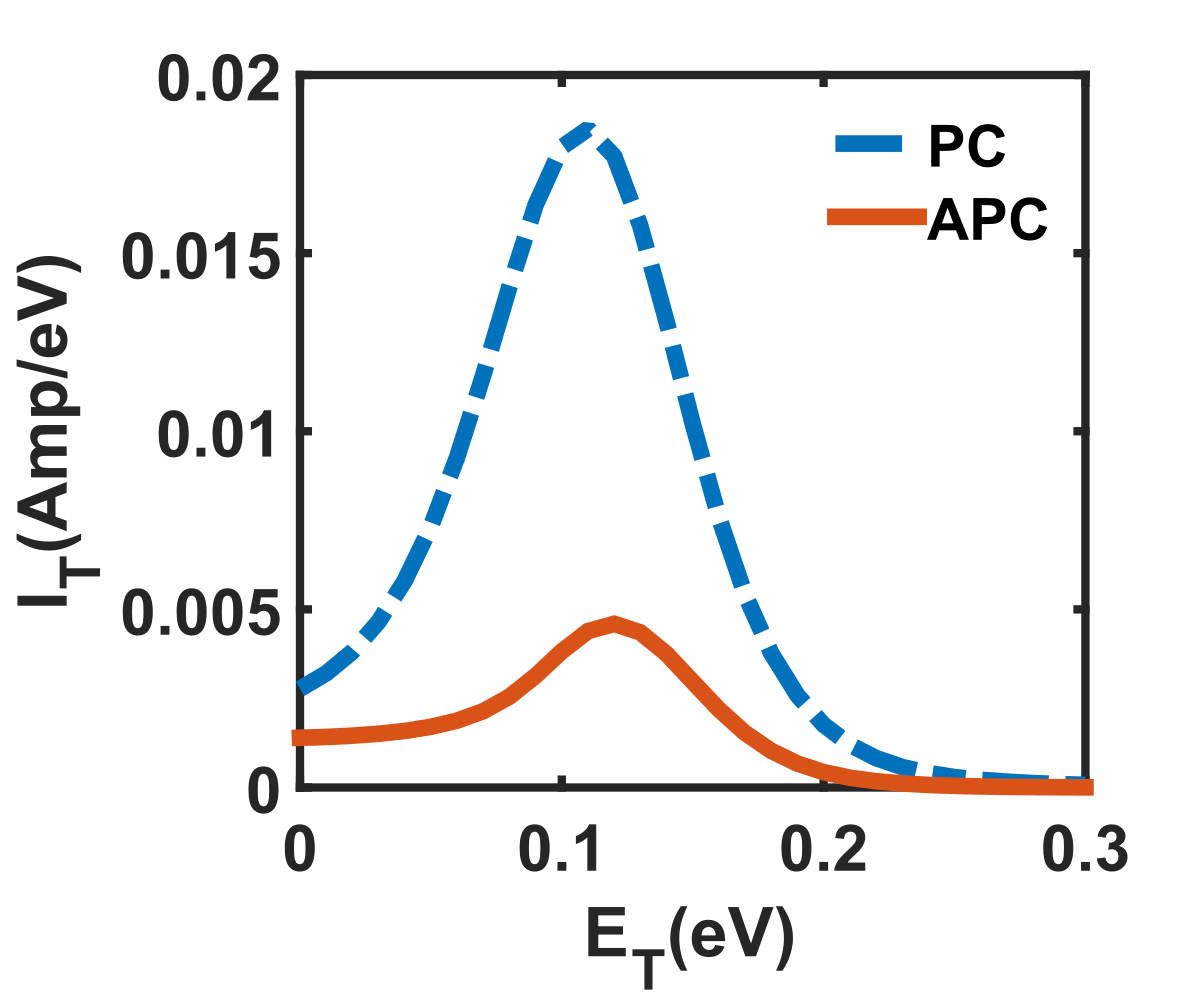}
	}\subfigure[]{\includegraphics[width=1.7in,height=1.25in]{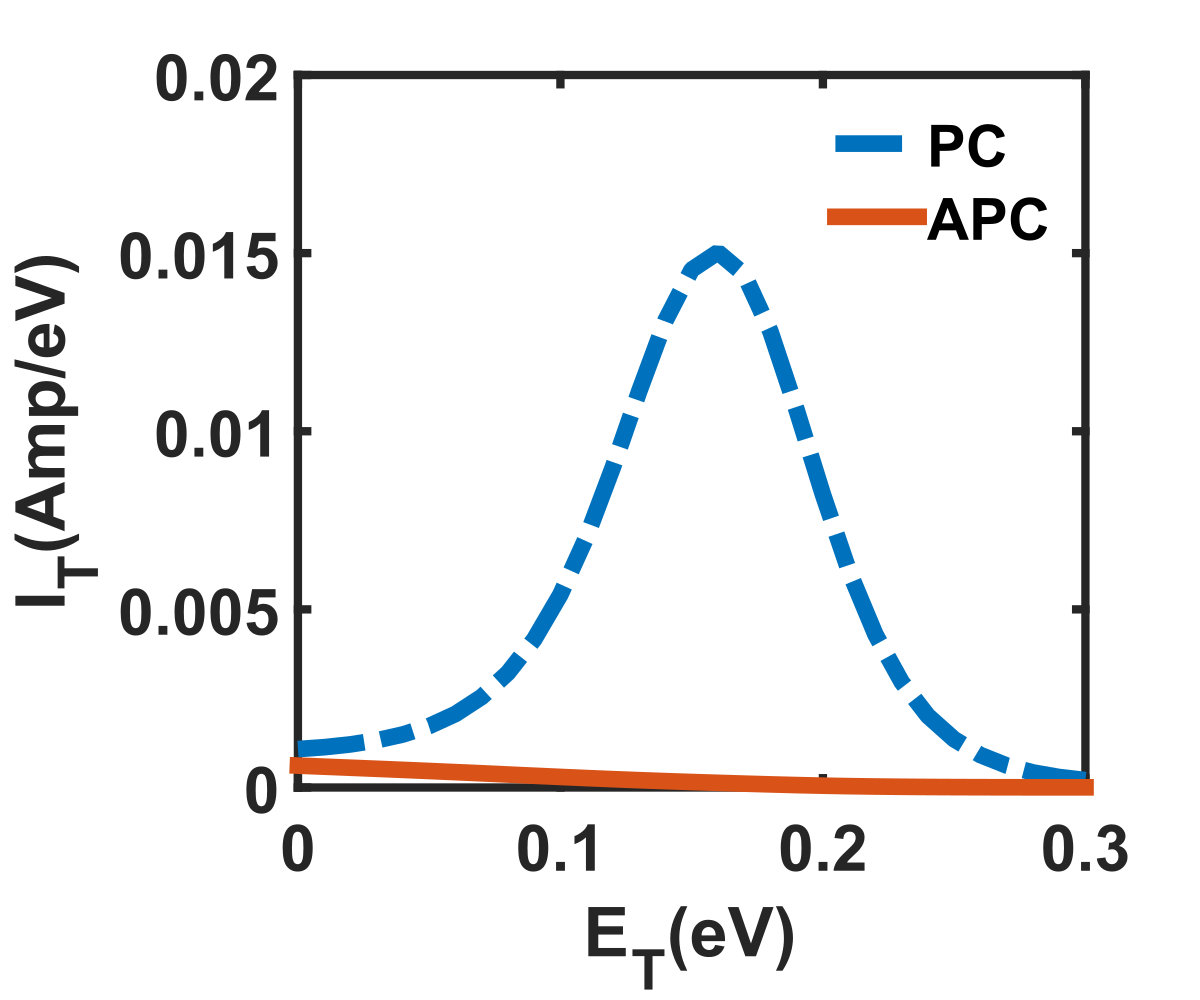}
}
\caption{Transverse mode profiles. Energy resolved charge current carried by per unit transverse mode energy for (a) the RTMTJ-I and (b) the RTMTJ-II device. The mode profile is calculated at fixed voltage of $0.05$V in both cases.}
\label{ModeProfile}
\end{figure}
\indent For the anti-parallel configuration sketched in Fig.\ref{band_diagram}(c) and (d), both the up and down-spin electrons do not have transmission peak resulting in a negligible current flow. This has been confirmed using NEGF simulation of the structure as shown in Fig.\ref{T_spectra}. \\
\begin{figure}[htb!]
	\subfigure[]{\includegraphics[scale=0.22]{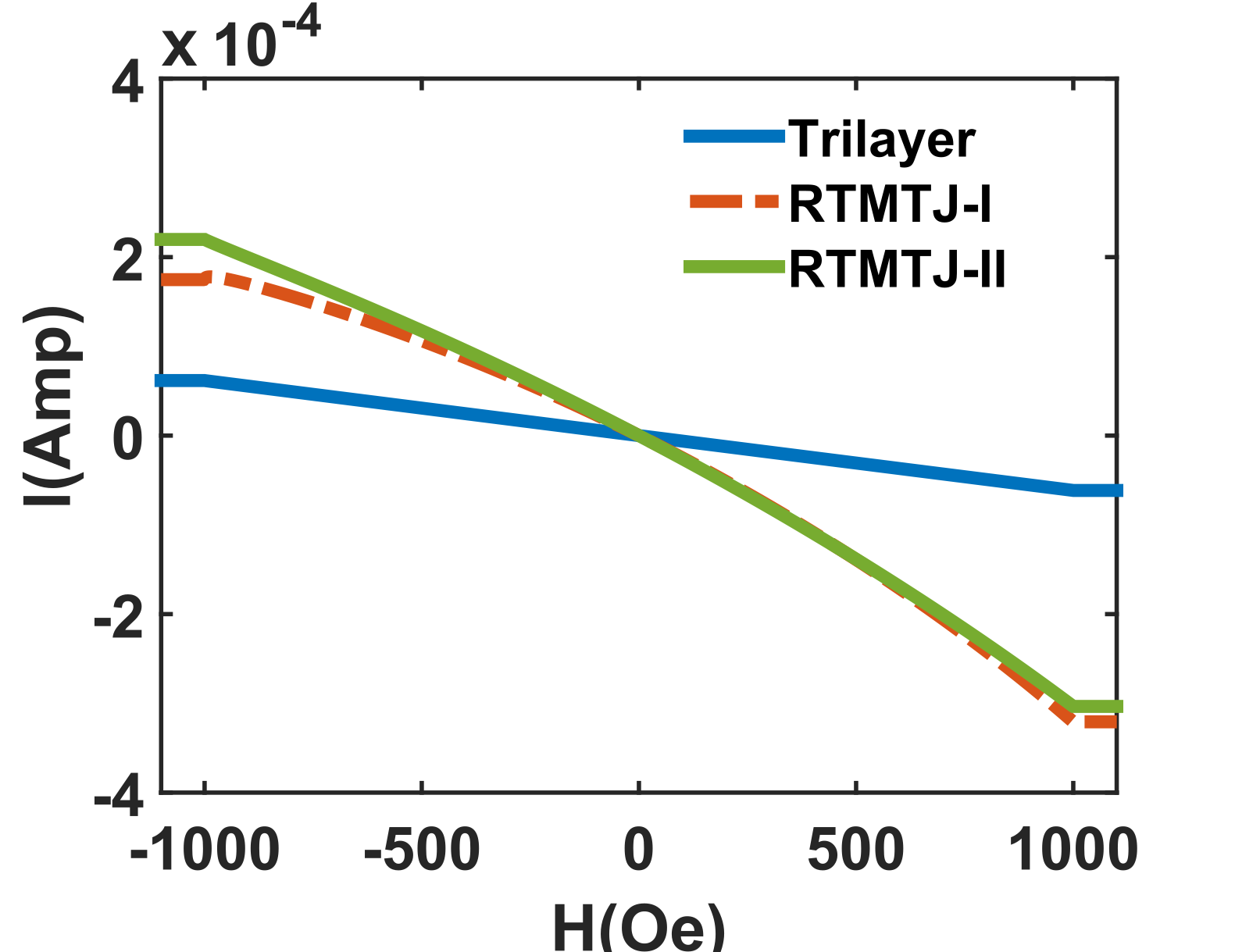}
	}\subfigure[]{\includegraphics[scale=0.22]{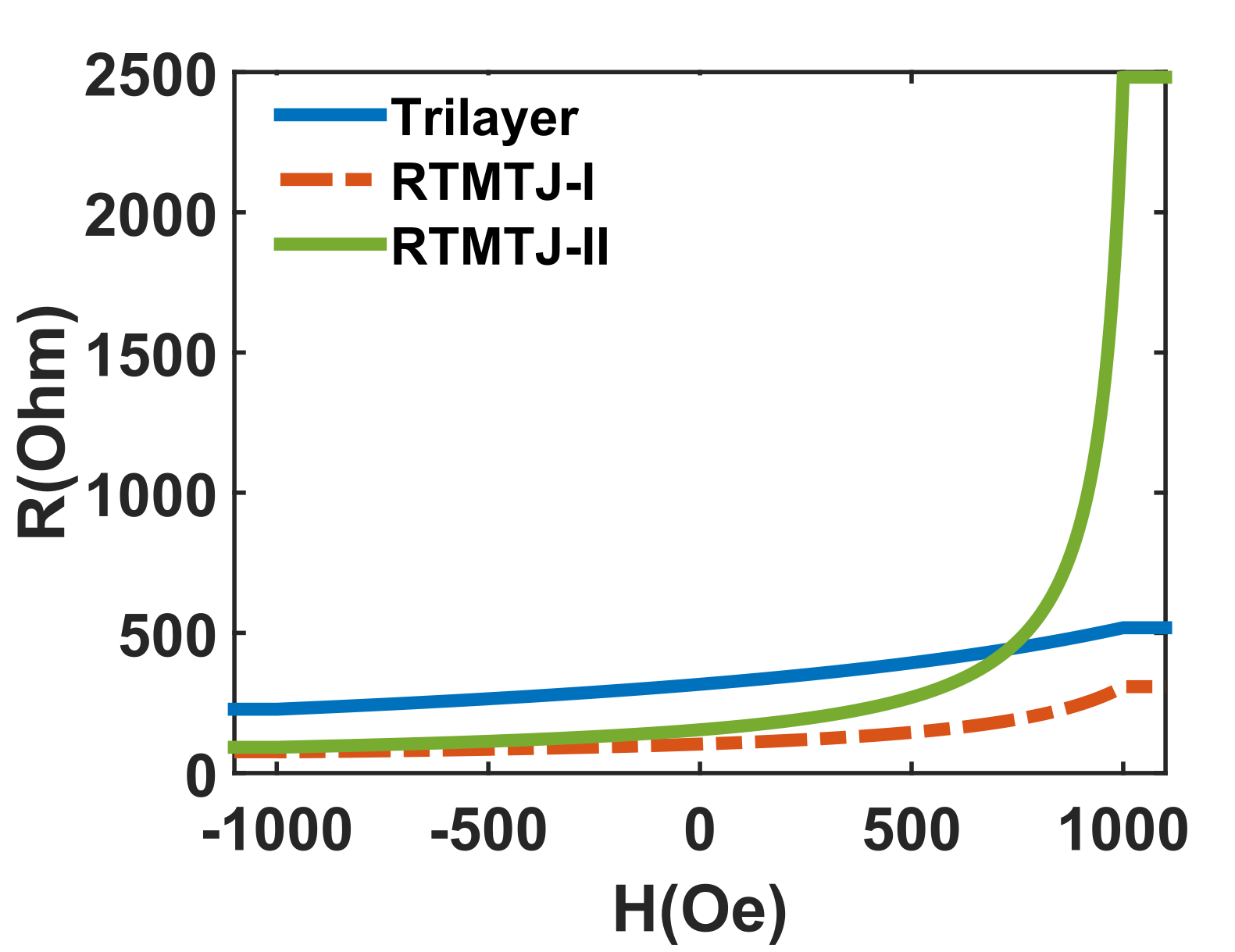}
}
\caption{Sensor performance with applied field. (a) Change in the current in a trilayer device (blue) and the RTMTJ devices (red and green) with the applied field ($H$).(b) Variation of device resistance with sensing field in a trilayer device (blue) and an RTMTJ devices I and II (red and green). The current and resistance both saturate at $H = H_k$(1000 Oe).}
\label{I_R_H}
\end{figure}
\indent The TMR of an RTMTJ device may be tuned by positioning the peak which is just below the exchange energy $\Delta$ so that a wide range of TMR values may be obtained. It is interesting to note from Fig.\ref{T_spectra}, that above exchange split energy $\Delta$, the position of the transmission peak for both the up and down-spin electrons is nearly the same for the parallel configuration and overlaps with each other for the anti-parallel configuration.  Hence, a high TMR in an RTMTJ device is not simply because of different positions of up- and down-spin transmission peaks as concluded in \cite{chatterji}, but rather due to the absence of one type of peak below the exchange split energy $\Delta$. \\
\indent Notably, the TMR of the RTMTJ design may vary by a large extent depending on the choice of the semiconductor heterojunction. The peak TMR values vary from a few thousands to a few hundreds \cite{chatterji}. Fig.\ref{ModeProfile} shows the mode profile of the charge current/unit energy versus the transverse mode energy. The transverse mode of conduction may be appropriately visualized as a set of parallel band diagrams with an offset equalling the transverse mode energy with respect to a fixed Fermi level. Total current carried by the device is given by area under mode profile. For the RTMTJ-I design, the total current carried in the parallel configuration and the anti-parallel configuration is comparable as shown in Fig.\ref{ModeProfile}(a), resulting in a lower TMR. For the RTMTJ-II design, the positions of the transmission peaks are such that there is a huge difference between total current carried in parallel and anti-parallel configurations,  as shown in Fig.\ref{ModeProfile}(b), leading to ultra high TMR, compared to the RTMTJ I design.  For the analysis to follow, we will use the two designs described above and compare them with the trilayer design.
\subsection{Sensor performance analysis} 
The functionality of a typical MR sensor is quantified via the following three parameters:
\begin{eqnarray}
S_I &=& \frac{dI(H)}{dH}, \label{Curr_Sens} \\
S &=& (dR/dH)/R(0), \label{Field_Sens} \\
NL &=&100\times\frac{R(H)-R_0-r_0H}{R(H)} , \label{NL}
\end{eqnarray}
\begin{figure}[tbh!]
	\subfigure[]{\includegraphics[scale=0.22]{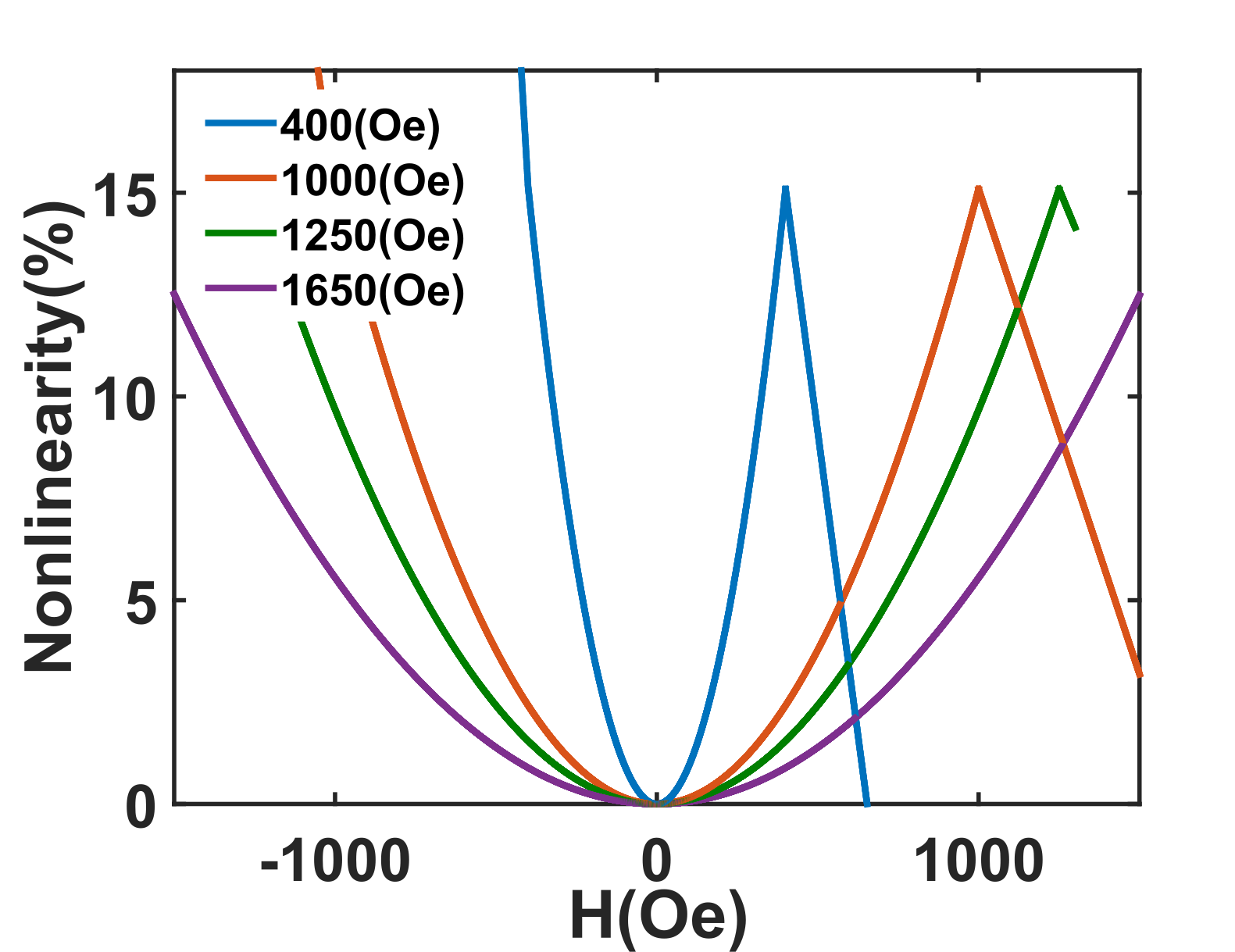}
	}\subfigure[]{\includegraphics[scale=0.22]{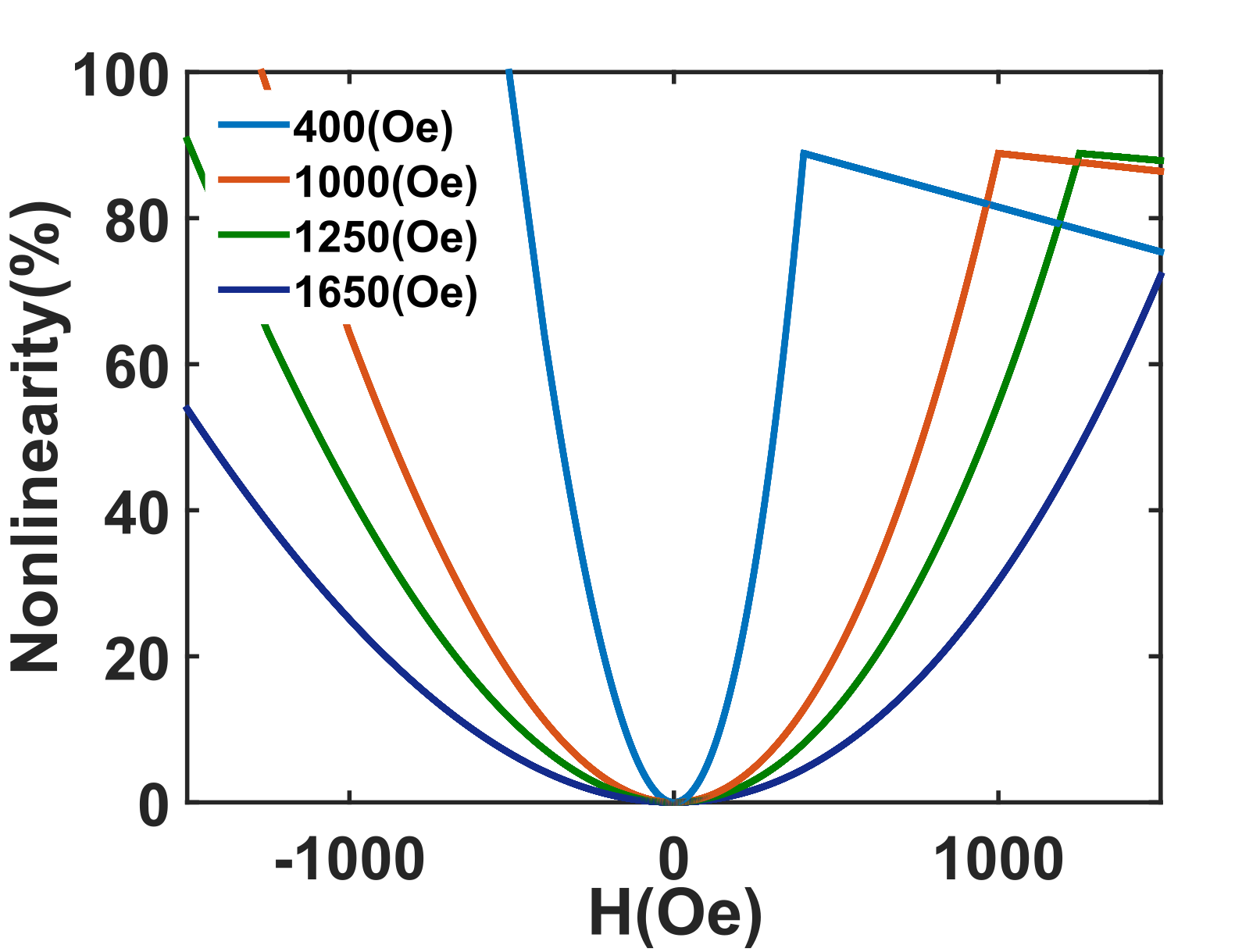}
}
%\vspace{.15in}
\caption{Nonlinearity variation with sensing field for various $H_k$ anisotropy fields (a) for the trilayer device (b) for the RTMTJ device II.}
\label{NL_H}
\end{figure}
where $S_I$, $S$, and $NL$ represent the current sensitivity, the field sensitivity or sensitivity and the non-linearity respectively. The first two parameters represent its sensitivity to current and magnetic field changes. Here, $I(H)$ is the current as a function of the applied magnetic field $H$, $R(H)$ and $R(0)$ represent the device resistance at a magnetic field of $H$ and at zero field respectively. The non-linearity parameter represents the percentage deviation of the actual resistance from a linear fit described by the slope $r_0$ and intercept $R_0$. This parameter must be typically kept below one percent to ensure a linear calibration at a given field. In order to properly quantify our structures, we need to calculate currents at different applied magnetic fields and hence different stable positions of the free magnet. \\
\indent We first examine in Fig.~\ref{I_R_H}, the change in the current and resistance of the three designs as a function of the sensing field. At small fields, we see that the change in current and resistance in all cases is linear and at higher fields $H > H_k$, the magnetization of the free layer aligns along the magnetic field, resulting in current saturation. Therefore, these devices are viable linear MR sensors for fields $H < H_k$. \\
\indent As seen in Fig.~\ref{I_R_H}(a), the slope of the change in current with respect to the sensing field in the RTMTJ case is $2.5\times10^{-4}$ mA/Oe greater compared to the trilayer MTJ device ($6.153\times10^{-5}$ mA/Oe). Thus, the current sensitivity is $306\%$ higher in the RTMTJ based sensor. Also, it can be inferred from Fig.~\ref{I_R_H}(b), that the trilayer resistance varies over a range $227\Omega$ to $518\Omega$, while that of the RTMTJ with changes from $76\Omega$ to $308\Omega$ for RTMTJ device I and between  $92\Omega$ to $2482\Omega$ for RTMTJ device II. 
This implies that the range over which the resistance of the RTMTJ device varies can be controlled by adjusting the device TMR \cite{chatterji}.\\
%\indent As seen in Fig.~\ref{I_R_H} the change in the current in the RTMTJ is much higher than the change in the current in the trilayer MTJ. The slope of the change in the current of the RTMTJ is $2.5\times10^{-4} mA/Oe$ mA/Oe compared to $6.153\times10^{-5}$ mA/Oe in the trilayer MTJ. Thus the current sensitivity ($S_I=\frac{dI(H)}{dH}$) of the RTMTJ is  $306\%$  higher.
\indent The non-linearity trends are depicted in Fig.~\ref{NL_H}, for different anisotropy fields $H_k$, translating to varying thicknesses of the sensing (free) layer \cite{VanDijken2005}. These trends follow from the assumption of coherent rotation of the magnetization of the free layer\cite{Slonczewski2009} and matches to a large extent with the experientially observed trilayer device trends\cite{VanDijken2005,zeng2012} . Also, in our simulations, we have assumed that the interlayer coupling between the two ferromagnets is absent which may introduce a small deviation along the $x$-axis in the $R-H$ and non-linearity curves. It can be inferred from Fig.~\ref{NL_H}(b) that RTMTJ devices have high non-linearity compared to trilayer devices for same value of anisotropy field.
\begin{figure}[thb!]
	\subfigure[]{\includegraphics[scale=0.23]{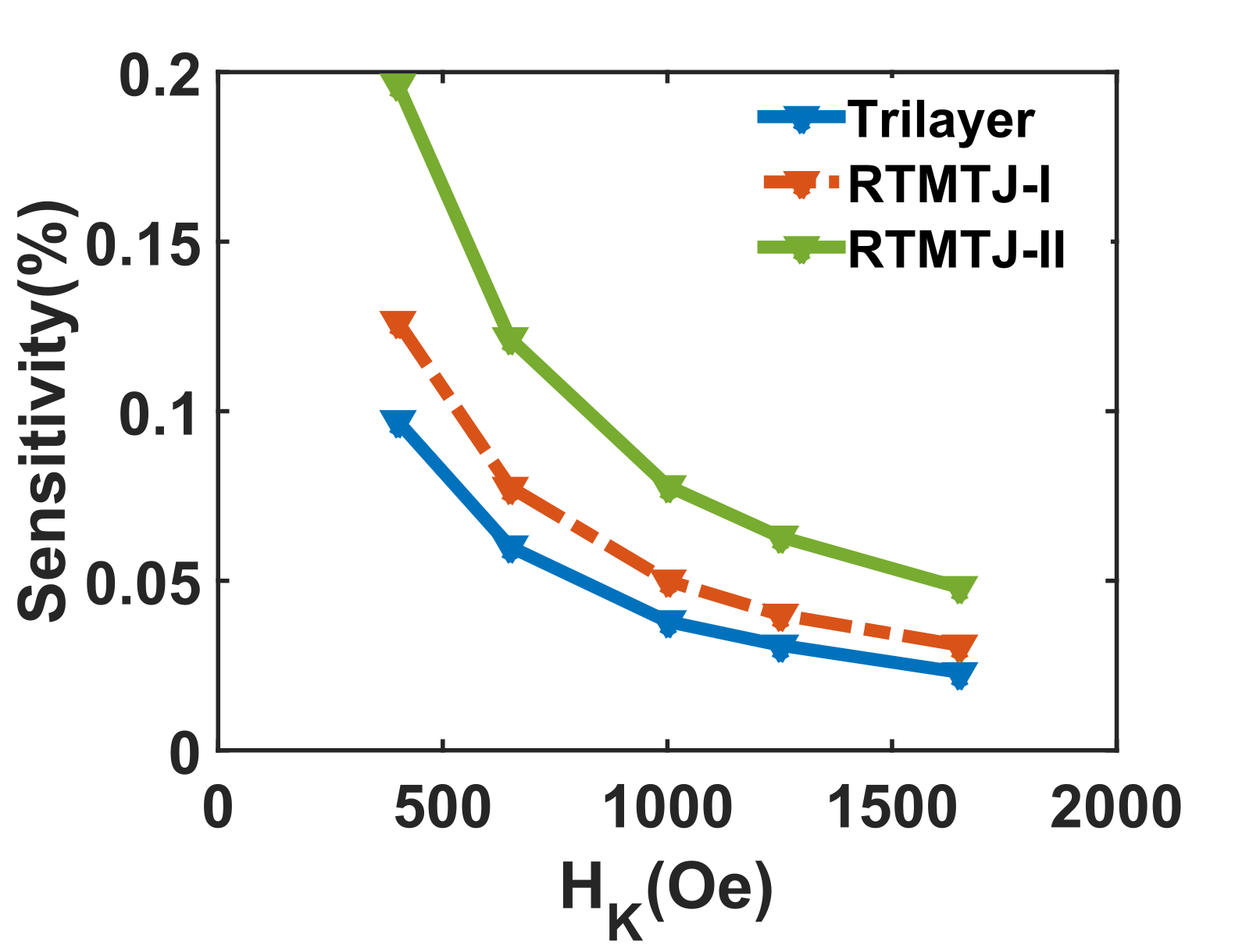}
	}\subfigure[]{\includegraphics[scale=0.23]{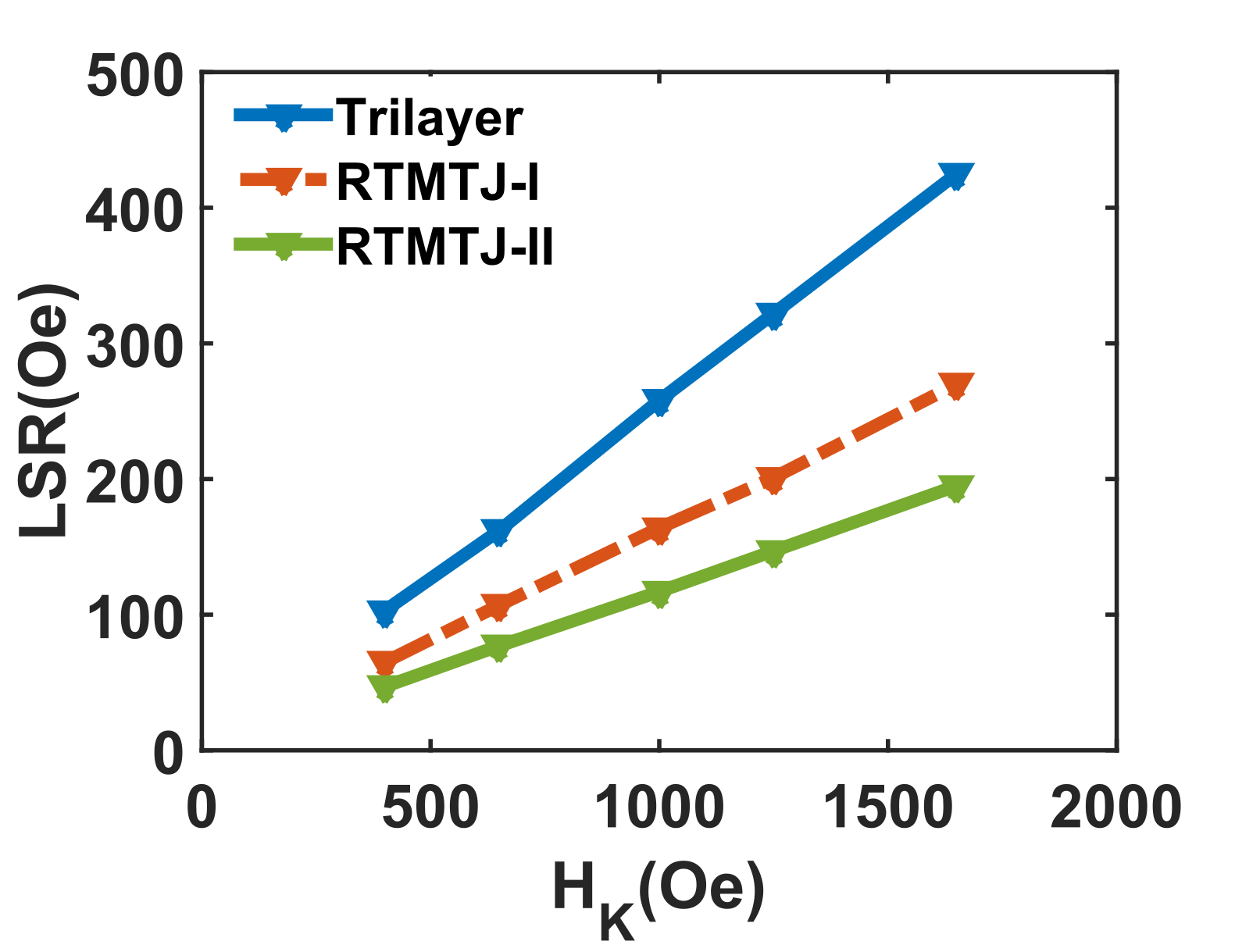}
}
%\vspace{.15in}
\caption{Variation of sensor performance with anisotropy field. (a) Variation in field sensitivity with anisotropy field. (b) Variation in the linear sensing range (LSR) with anisotropy Field. The LSR is defined as the range over which $NL<1\%$.}
\label{S_LSR_H}
\end{figure}

\indent We note in Fig.~\ref{S_LSR_H}, that for all devices, the field sensitivity \cite{zeng2012} decreases with increasing $H_k$, and the linear field range widens with increasing $H_k$. Notably, for the trilayer device, we have obtained $S=0.02\%$ and a sensing range of $-425<H(Oe)<425$ for an anisotropy field $H_k=1650$Oe, corresponding to a sensing layer thickness of $t=1.36$nm \cite{zeng2012}. It can be inferred from Fig.~\ref{S_LSR_H}(a), that the RTMTJ device is $105\%$ more sensitive than the trilayer device. The sensitivity of the RTMTJ structure can be increased without decreasing the anisotropy field, which potentially results in a higher noise immunity in comparison with trilayer devices\cite{Garcia-Palacios1998}.
In the RTMTJ devices, the availability of a wider design landscape \cite{chatterji} catered toward a target TMR, enables one to generate a family of curves for the field sensitivity and linear sensing range as shown in Fig.~\ref{S_LSR_H} for the two TMR cases.
\begin{figure}[tb!]
	\subfigure[]{\includegraphics[width=1.8in]{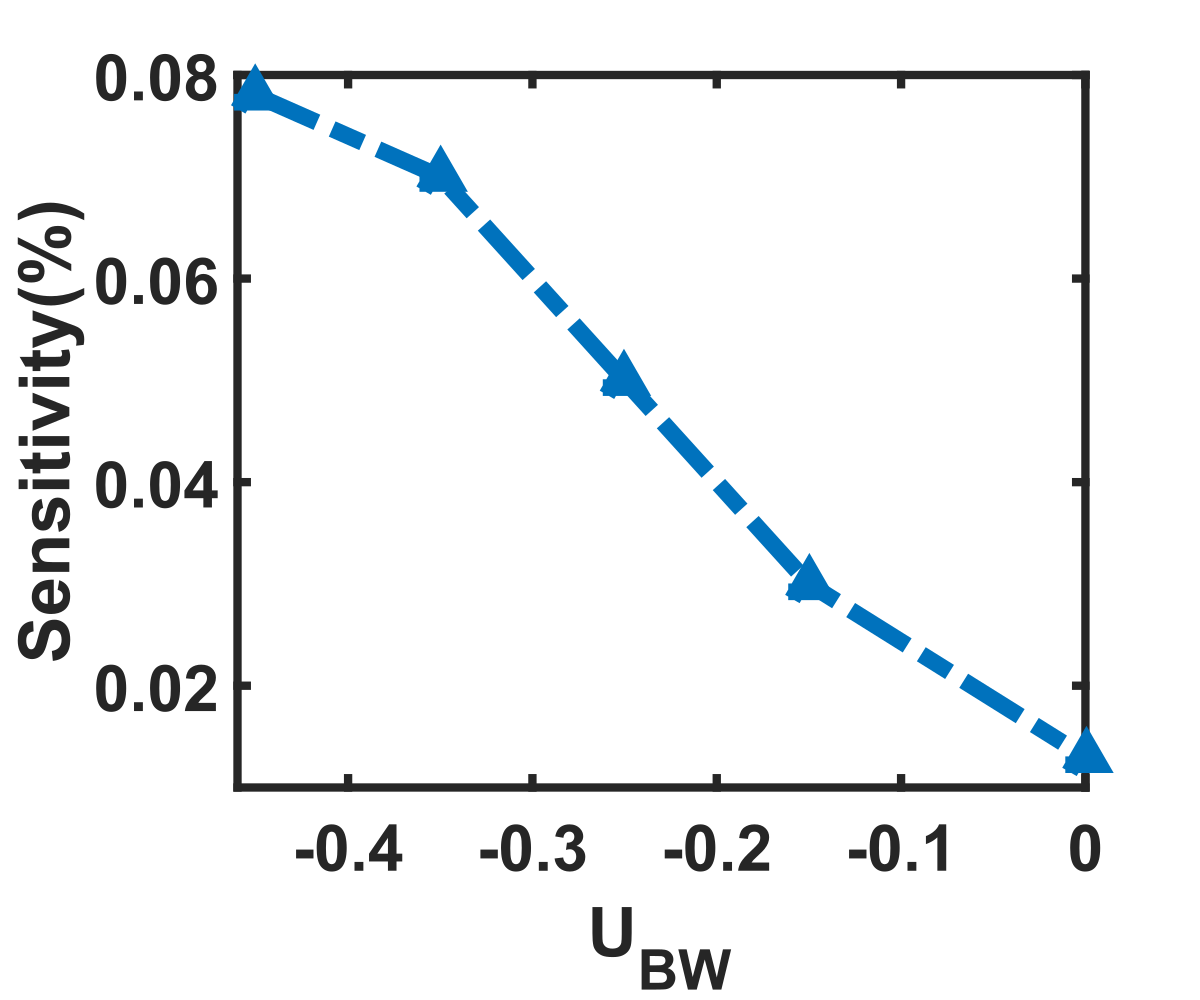}	}\subfigure[]{\includegraphics[width=1.8in]{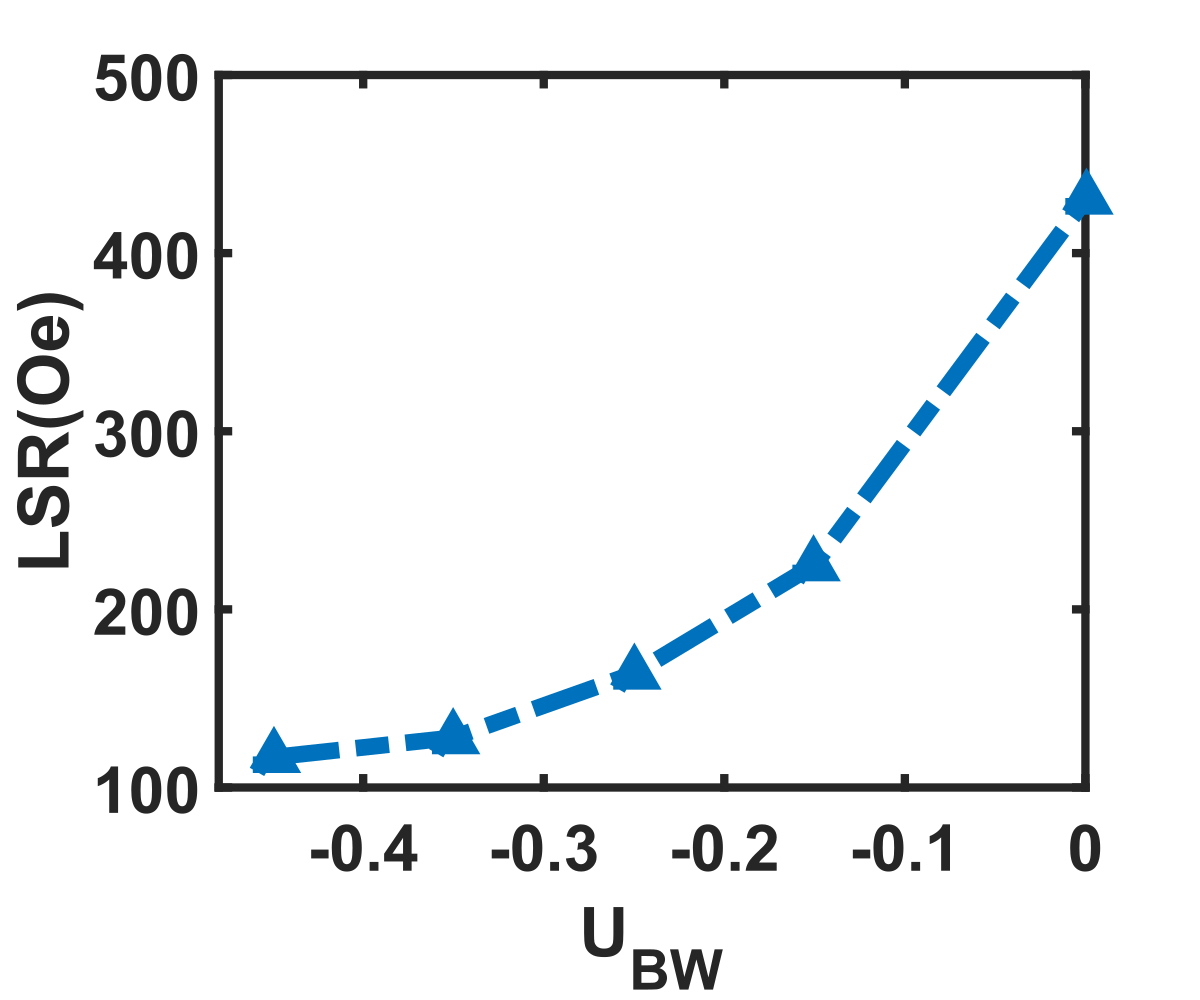}}
	\caption{Sensor performance variation with band-profile. (a) Sensitivity variation with difference between bottom of conduction band of ferromagnet and semiconductor $U_{BW}$ (b) Variation of Linear sensing range(LSR) with $U_{BW}$. Thickness of semiconductor $W_{W}=1$nm and Anisotropy field $H_{K}=1000Oe$ kept fixed for all the devices during the variation of $U_{BW}$.}
	\label{Sh_LSR_Vs_Ubw}
\end{figure}

\begin{figure}[htb!]
	\subfigure[]{\includegraphics[width=1.83in]{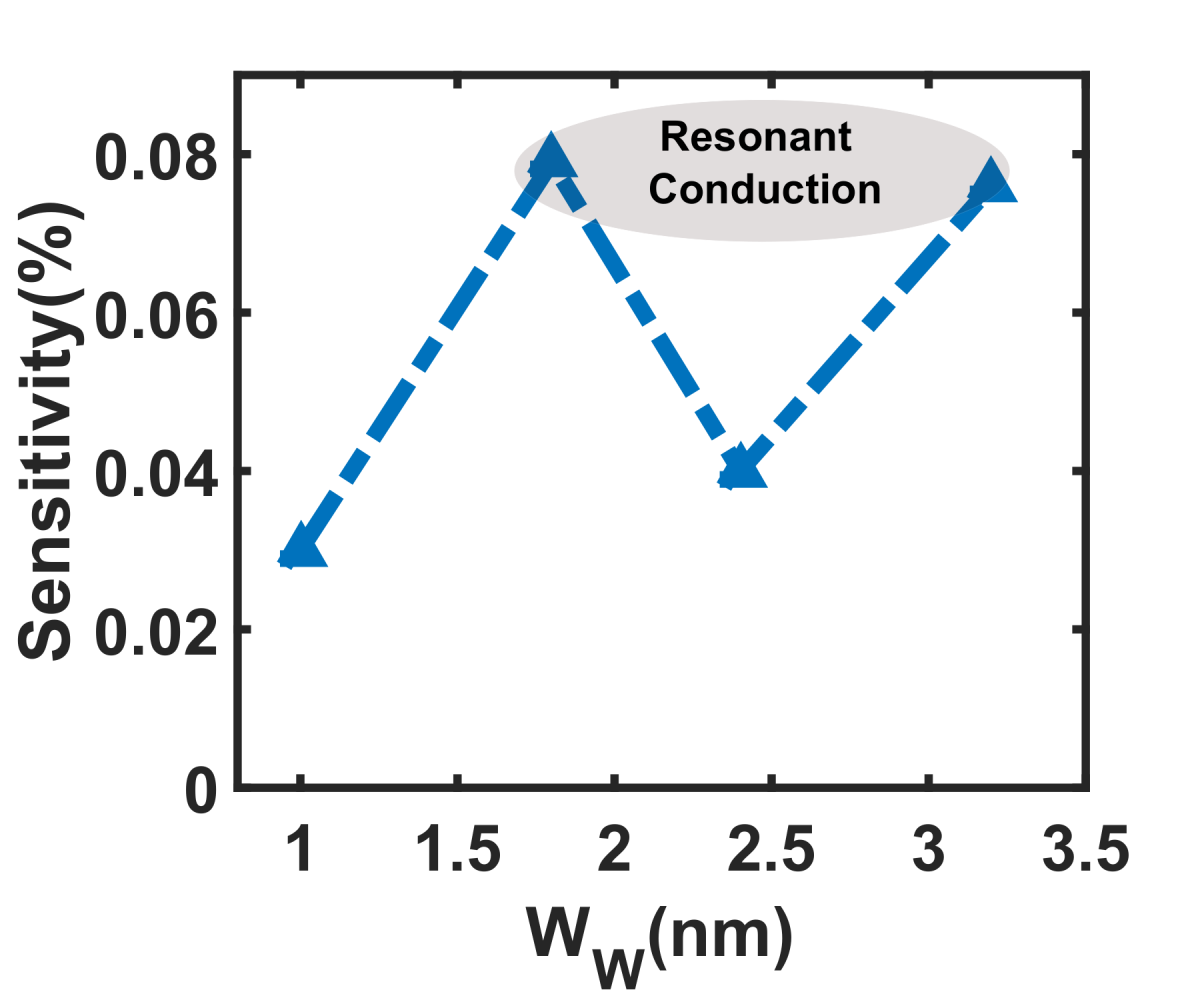}	}\subfigure[]{\includegraphics[width=1.83in]{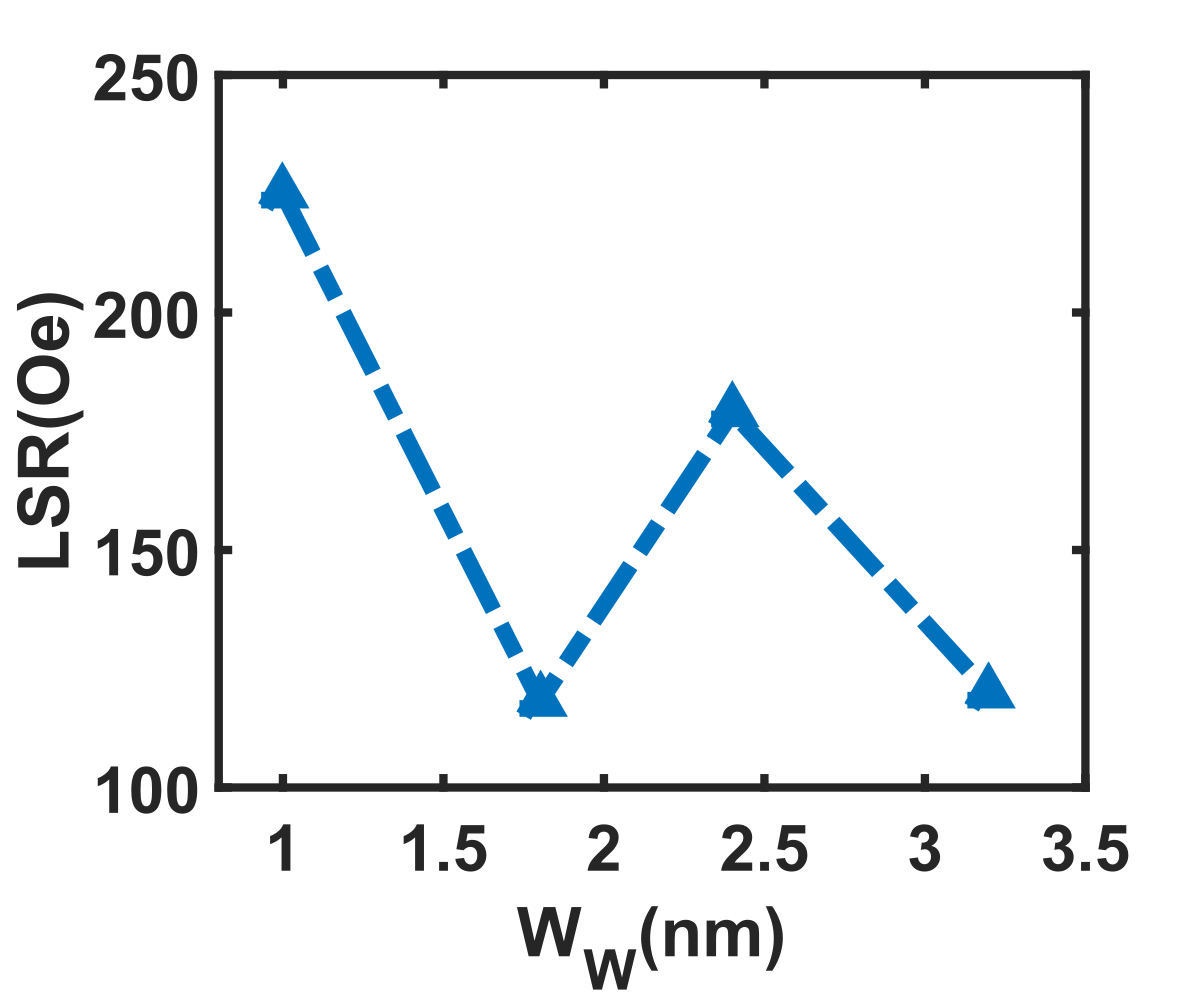}}
	\caption{Sensor performance variation with structural modulation. (a) Sensitivity variation with semiconducting layer thickness $W_{W}$ (b) Variation of Linear sensing range(LSR) with $W_{W}$. $U_{BW}$ and anisotropic field is kept fixed at $-0.15eV$ and $1000Oe$ respectively for all the devices during the variation of $W_{W}$.}
	\label{Sh_LSR_Ww}
\end{figure}
\subsection{Structural impact on performance}
The utility of the RTMTJ structure becomes apparent with its ability to span a large range of sensor performance parameters depending on the application. The positions of transmission peaks provide highly sensitive tuning knobs to vary the TMR of the device, which in turn, modulates its sensitivity and linear sensing range. There are two primary device parameters that vary the position of the transmission peaks, namely the barrier height, $U_{BW}$, and the thickness of semiconductor layer, $W_W$. We notice from Fig.\ref{Sh_LSR_Vs_Ubw} that the sensitivity and LSR vary monotonically with $U_{BW}$.
The sensitivity and LSR variation as a function of the thickness of the semiconductor layer, $W_W$ are shown in Fig.\ref{Sh_LSR_Ww}. The position of transmission peaks in well region, however, vary non-monotonically with varying thickness of SC layer resulting in the sensitivity and LSR plots as noticed in Fig.\ref{Sh_LSR_Ww} . Specifically, it can seen from Fig.\ref{Sh_LSR_Ww}(a) that when resonant conduction occurs, ($W_W=1.8$nm and $ W=3.2$nm), a field sensitivity which is $105\%$ higher than the trilayer device is obtained. We can thus infer that the RTMTJ indeed offers a wide functional selectivity via simple structural variation.
\subsection{Optimal sensor design}
So far, we have focussed on the traditional design setting in which the sensing layer is kept out of the plane at a right angle to the fixed layer under equilibrium conditions. 
\begin{figure}
	\centering
	\includegraphics[width=3.5in]{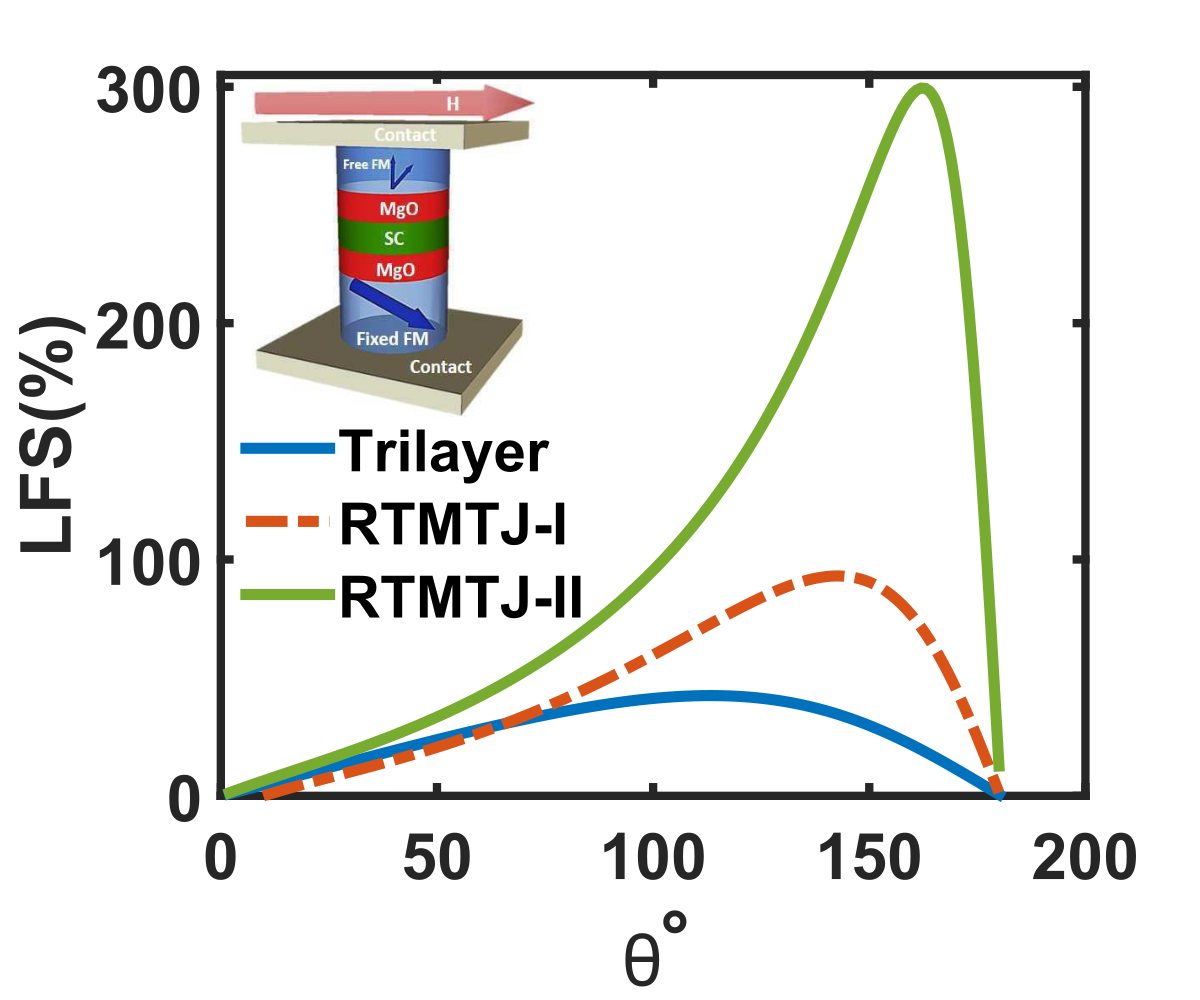}
	\caption{Finding the optimal operating angle: Local field sensitivity (LFS) variation with the angle between the fixed and free layers of the trilayer (blue), the RTMTJ-I (red) and the RTMTJ-II (green) devices. The inset depicts a schematic of the proposed device tailored for a ultra-high sensitivity.}
	\label{LocalSensitivity}
\end{figure}
Here, we consider a case when such an design criteria be relaxed and propose a different sensor design which can be used to obtain ultra-high sensitivity without much change in the LSR. To explore such a design possibility we have extended the definition of field sensitivity to the local (angular) field sensitivity (LFS) defined as:
\begin{eqnarray}
LFS &=& \frac{1}{R(\theta)}\frac{dR}{d\theta}. \label{Local_Field_Sens},
\end{eqnarray}
where $R(\theta)$ is the resistance of the device when the free and fixed layers are kept at an equilibrium angle of $\theta$. We notice from Fig.\ref{LocalSensitivity}, that the RTMTJ sensors have a pronounced LFS at $\theta=144^{\circ}$ and $\theta=163^{\circ}$. This pronounced behavior can be used to tap the potential of the RTMTJ based sensor to exhibit an ultra-high sensitivity. Such an orientation of magnetization such that the fixed layer is out of a plane can be fixed during fabrication via magnetic field annealing \cite{Grady,Tsunoda}, resulting in the fixed and free layers aligned at an azimuthal angle $\theta$ at equilibrium.  
\begin{table}[]
	\centering
	\label{tableS}
		\caption{Sensitivity and linear sensing range (LSR) for RTMTJ sensor in comparison with the trilayer device. The anisotropy field is kept constant at $1000Oe$ for all the devices.}
	\begin{tabular}{|l|r|r|l|r|}
		\hline
		Device   & $\theta^{\circ}$ & Sensitivity & LSR(Oe)                        & \begin{tabular}[c]{@{}r@{}}\% Increase in \\ sensitivity\end{tabular} \\ \hline
		Trilayer      & 90$^{\circ}$     & 0.038\%     & \multicolumn{1}{r|}{258Oe}     & 0                                                                     \\ \hline
		RTMTJ-I  & 144$^{\circ}$    & 0.091\%      & -150--200Oe                    & 136\%                                                                 \\ \hline
		RTMTJ-II & 163$^{\circ}$    & 0.290\%      & \multicolumn{1}{r|}{-47--55Oe} & 663\%                                                                 \\ \hline
	\end{tabular}
\end{table}

It can be inferred from the Tab.I that the optimum design presented may be used to further enhance the sensitivity of the RTMTJ based structure.
\section{Conclusion}
We have demonstrated that the non-trivial spin filtering physics accompanying double barrier resonant tunneling can lead to an ultra-high TMR, which can be sensitively tuned. Using this, we presented MR based magnetic field sensor device designs featuring exceptional current sensitivity and linear sensing range. In particular, the current sensitivity of such a device was shown to be $300\%$ higher than that of a typical trilayer device. The RTMTJ device designs also offer a much better design flexibility in comparison to trilayer devices. Using the angle dependent spin filtering physics, we demonstrated an optimal sensor design whose sensitivity can be further enhanced to around $700\%$. It is thus envisioned that the high sensitivity of RTMTJ device might also be used for pico-Tesla magnetic field sensors \cite{Liou2011}, by making suitable design modifications. Our work thus paves new directions in exploring spintronic device functionalities to be tapped via engineering novel spin filtering paradigms \cite{Butler2001,Bauer_TSTT}. \\\\
{\it{Acknowledgements:}} It is our pleasure to acknowledge insightful discussions with Prof. Supriyo Datta. The authors acknowledge Niladri Chatterji for his initial contributions and useful discussions. This work was in part supported by the IIT Bombay SEED grant and the Department of Science and Technology (DST), India, under the Science and Engineering Board grant no. SERB/F/3370/2013-2014.

\bibliographystyle{apsrev}
\bibliography{refSTT}

\end{document}